\documentclass[12pt]{article}
\pdfoutput=1 
\usepackage{amsmath,amssymb}
\usepackage{graphicx}
\usepackage{color}
\usepackage[colorlinks]{hyperref}
\hypersetup{
    colorlinks=true,
    linkcolor=black,
    citecolor=black,
    filecolor=black,
    urlcolor=black,
    bookmarks=true,
    bookmarksopen=true,
    bookmarksnumbered=true,
}
\usepackage{enumitem}
\usepackage{youngtab}

\numberwithin{equation}{section}

\newcommand{\bea}{\begin{eqnarray}}
\newcommand{\eea}{\end{eqnarray}}
\def\no{\nonumber}
\newcommand{\half}{\frac{1}{2}}

\newcommand{\p}{\partial}

\newcommand{\tr}{\textrm{Tr}}
\newcommand{\Li}{\textrm{Li}}
\newcommand{\id}{1}
\newcommand{\ord}{{\cal O}}

\newcommand{\bC}{\ensuremath{\mathbb{C}}}

\newcommand{\bR}{\ensuremath{\mathbb{R}}}

\newcommand{\bZ}{\ensuremath{\mathbb{Z}}}

\newcommand{\scN}{\ensuremath{\mathcal{N}}}

\newcommand{\scP}{\ensuremath{\mathcal{P}}}

\newcommand{\scW}{\ensuremath{\mathcal{W}}}

\newcommand{\be}{\begin{equation}}
\newcommand{\ee}{\end{equation}}
\newcommand{\beq}{\begin{eqnarray}}
\newcommand{\eeq}{\end{eqnarray}}

\newcommand{\bara}{\ensuremath{\overline{1}}}
\newcommand{\barb}{\ensuremath{\overline{2}}}
\newcommand{\ybullet}{\ensuremath{\bullet}}
\newcommand{\lone}{\ensuremath{l_1}}
\newcommand{\ltwo}{\ensuremath{l_2}}
\newcommand{\lthree}{\ensuremath{l_3}}
\newcommand{\lm}{\ensuremath{l_m}}

\begin{document}

\begin{titlepage}
\begin{flushright}
\normalsize
Imperial/TP/2012/JE/03 \\
LPTENS-12/44 \\
PUPT-2430
\end{flushright}

\vfil

\bigskip

\begin{center}
\LARGE
Wilson Loops in 5d $\mathcal{N}=1$ SCFTs \\
and AdS/CFT
\end{center}

\vfil
\medskip

\begin{center}
\def\thefootnote{\fnsymbol{footnote}}

Benjamin Assel$^{\natural}$, John Estes$^\flat$ and Masahito Yamazaki$^\sharp$

\end{center}

\begin{flushleft}\small

\vskip 5mm

\centerline{$^\natural$ Laboratoire de Physique Th\'eorique de l'\'Ecole Normale Sup\'erieure, }
\centerline{24 rue Lhomond, 75231 Paris cedex, France}

\vskip 5mm

\centerline{$^\flat$ Blackett Laboratory, Imperial College,}
\centerline{London, SW7 2AZ, United Kingdom}

\vskip 5mm

\centerline{$^\sharp$
Princeton Center for Theoretical Science, Princeton University, }
\centerline{Princeton, NJ 08544, USA}
\end{flushleft}


\thispagestyle{empty}

\setcounter{tocdepth}{2}


\begin{center}
{\bfseries Abstract}
\end{center}

We consider $\half$-BPS circular Wilson loops in a class of 5d
 superconformal field theories on $S^5$.
The large $N$ limit of the vacuum expectation values of Wilson loops
are computed both
by localization in the field theory and by evaluating the fundamental string
and D4-brane actions in the dual massive IIA supergravity background.
We find agreement in the leading large $N$ limit
for a rather general class of representations, including
fundamental, anti-symmetric and symmetric representations.
For single node theories the match is straightforward, while for quiver
 theories, the Wilson loop can be in different representations for each
 node.
We highlight the two special cases when the Wilson loop is in either in all symmetric or all anti-symmetric representations.  In the anti-symmetric case, we find that the vacuum expectation value factorizes into distinct contributions from each quiver node.  In the dual supergravity description, this corresponds to probe D4-branes wrapping internal $S^3$ cycles.
The story is more complicated in the symmetric case and the vacuum expectation value does not exhibit factorization.

\end{titlepage}

\newpage
\setcounter{page}{1}

\tableofcontents

\section{Introduction}

Wilson loops are important gauge-invariant observables in
gauge theories, and provide valuable dynamical information
of the system.
Since the pioneering works of
\cite{Rey:1998ik,Maldacena:1998im},
they have been studied extensively in the context of the AdS/CFT
correspondence.

In this paper, we consider Wilson loops in a class of 5d $\scN=1$
superconformal
field theories (SCFTs)
and their holographic duals. There are very few quantitative statements
on such Wilson loops in the literature 
(see however \cite{Young:2011aa} where the holography of non-BPS Wilson loops in 5d maximally supersymmetric SYM are considered).
Part of the reasons for this is that
5d gauge theories are non-renormalizable.  There is a danger that infinitely many irrelevant operators could
potentially contribute near the strongly coupled UV fixed point,
hence invalidating the computation from the effective Lagrangian.
The goal of this paper is to overcome this difficulty, by first computing the vacuum expectation value of the Wilson loops in the effective theory at strong coupling using localization techniques and then comparing to the dual supergravity description, which provides a definition of the strongly coupled UV fixed point.

We consider a class of 5d $\scN=1$ SCFTs discovered in
\cite{Intriligator:1997pq} (see also
\cite{Morrison:1996xf,Seiberg:1996bd})  and generalized recently to
quiver theories
in \cite{Bergman:2012kr}. These theories are dual to warped $AdS_6\times
S^4/\bZ_n$ compactifications in massive type IIA supergravity
\cite{Brandhuber:1999np,Bergman:2012kr} (for massive IIA supergravity see \cite{Romans:1985tz}
and \cite{Lozano:2012au} for recent T-dual type IIB backgrounds
), and are engineered from type I'
string theory on $\bR^{4,1}\times \bC^2/\bZ_n \times \bR$
with $N$ D4-branes, $N_f$ D8-branes and one O8$^-$-plane.
These 5d $\scN=1$ theories are specified by the choice of $N, N_f$ and $n$.\footnote{When $n$ is even, there is
an extra two-fold choice, corresponding
to a compactification with or
without vector structure. However, as we will see the two choices give identical
Wilson loop VEVs in the leading large $N$ limit.} The existence of the fixed point requires
$N_f< 8$ \cite{Seiberg:1996bd};
this is a necessary condition for the inverse square effective gauge
coupling constant to stay positive everywhere on the Coulomb branch of
the moduli space. In this case the moduli space is smooth and we could take the strong coupling
limit where the bare
gauge coupling constant goes to infinity.  One can then argue, without proof, for the existence of the UV fixed point at the intersection (origin) of the Coulomb and Higgs branches.

We consider these 5d SCFTs on the Euclidean $S^5$. We compute the vacuum expectation value (VEV) of the $\half$-BPS circular Wilson line operator,
placed on the great circle of $S^5$:
\beq
\langle W_R \rangle = \frac{1}{\textrm{dim} R} \Big\langle \textrm{Tr}_R \,
\mathcal{P}  \exp \int ( i A_{\mu} \dot{x}^{\mu} + \sigma \dot{y})
\Big\rangle \ ,
\label{loopdef}
\eeq
where $A_{\mu}$ is the gauge field, $\sigma$ is the scalar in the 5d
vector multiplet, $\scP$ a path-ordered product, and
$R$ a representation of the gauge group.
We also wrote the worldline of the Wilson line as $x^{\mu}(\tau)$,
parametrized by $\tau$, and introduced the function $y(\tau)$, which can
be thought of as a path in the internal space.
$\half$-BPS supersymmetry requires $|\dot{x}|^2=\dot{y}^2$.

We compute the VEV of this Wilson loop in a
general representation represented by a Young diagram,
both in gauge theory and gravity in the large $N$ limit.\footnote{For
similar computations for $\half$-BPS circular Wilson loops in 4d $\mathcal{N}=4$ theories,
see \cite{Yamaguchi:2006tq,Hartnoll:2006is,Gomis:2006sb}.}
The expression is simpler when we consider
\setlist{nolistsep}
\begin{enumerate}[label=({\arabic*})]
\item fundamental representation
\item $k$-th antisymmetric representation $A_k$, i.e., the anti-symmetric
      part of the $k$-th tensor product of the fundamental representation
\item $k$-th symmetric representation $S_k$, i.e. the symmetric part of the
      $k$-th tensor product of the fundamental representation
\end{enumerate}
of the gauge groups.

For the case $n=1$, we find a complete agreement in the leading large $N$ limit, with the VEVs given as:
\begin{align}
 \langle W_{\rm fund} \rangle & \sim \exp \left[6\pi
 \sqrt{\frac{N}{2(8-N_f)}} \right] \label{agreementf} \ , \\
 \langle W_{A_k} \rangle &  = \langle W_{A_{2N-k}} \rangle \sim \exp \left[4 \pi
 \sqrt{\frac{N^3}{2(8-N_f)}}\left(1-\left|1-\frac{k}{N}\right|^{3/2}\right)
 \right] \label{agreementa} \ , \\
 \langle W_{S_k} \rangle & \sim \exp \left[9 \pi
 \sqrt{\frac{N^{3}}{2(8-N_f)}}\left(\left(1+\frac{4k}{9N}\right)^{3/2}-1\right)
 \right] \label{agreements} \  .
\end{align}
As expected, we
find that for anti-symmetric representations $k$ is bounded, while $k$
can take arbitrary values for symmetric representations.  Additionally, $A_k$ is a reducible representation and in the leading large $N$ limit, where $k$ scales with $N$, only the largest irreducible representation gives the leading expression.  In the case $k$ is held fixed in the large $N$ limit, all of the expressions reduce to a product of fundamental Wilson loops.

We also discuss more general representations.
A representation of $USp(2N)$ is specified by a Young diagram with
at most $N$ rows.
When we have a Wilson line in the representation
specified by a partition $(k_1, \ldots, k_m)$, with $m$ held fixed in the large $N$ limit,
we find
\begin{align}
 \langle W_{(k_1, \ldots, k_m)} \rangle & \sim \exp \left[9 \pi
 \sqrt{\frac{N^{3}}{2(8-N_f)}}\sum_{i=1}^m
 \left(\left(1+\frac{4k_i}{9N}\right)^{3/2}-1\right)
 \right] \  .
 \label{generals}
\end{align}
Similarly, when we have a Wilson line in the representation
specified by a {\it dual} partition $(l_1, \ldots, l_{m})$, again with $m$ held fixed in the large $N$ limit, we find
\begin{align}
 \langle W_{(l_1, \ldots, l_{m})^T} \rangle & \sim \exp \left[4 \pi
 \sqrt{\frac{N^3}{2(8-N_f)}}\sum_{i=1}^{m} \left(1-\left|1-\frac{l_i}{N}\right|^{3/2}\right)
 \right] \ .
 \label{generala}
\end{align}
In both cases this leads to factorized expressions in the leading large $N$ limit:
\beq
\langle W_{(k_1, \ldots, k_m)}\rangle = \prod_i\, \langle W_{S_{k_i}} \rangle,
\quad
\langle W_{(l_1, \ldots, l_m)^T}\rangle = \prod_i\, \langle W_{A_{l_i}}
\rangle \ .
\eeq
Note that when the $k_i$ or $l_i$ are taken to be finite in the large $N$ limit, both expressions reduce to a product of fundamental representations, $\langle W_{\rm fund} \rangle$, at leading order.  In particular, this is consistent with self-dual partitions.  The cases where $m$ also scales with $N$ require the back-reaction of the D4-branes to be taken into account, along the lines of \cite{Lunin:2006xr,D'Hoker:2007fq}.

For $n>1$, the theories are linear quiver theories given by a total of $q$ products of $USp(2N)$ and $SU(2N)$ gauge factors.  For odd $n$, we have $q =[n/2]+1$ and gauge group $USp(2N) \times SU(2N)^{q-1}$.  For even $n$ with no vector structure, we have $q = [n/2]$ with gauge group $SU(2N)^q$.  For even $n$ with vector structure, we have $q = [n/2]+1$ with gauge group $USp(2N) \times SU(2N)^{q-2} \times USp(2N)$.  In either of the cases, the expressions generalize as follows.  When the Wilson loop is in the fundamental representation of a single node we have
\begin{align}
\label{fundorb}
\langle W_{\rm fund} \rangle & \sim \exp \left[6\pi
 \sqrt{\frac{nN}{2(8-N_f)}} \right] \ .
\end{align}
For arbitrary configurations of anti-symmetric representations, we find that the result factorizes into contributions from each node in the quiver
\begin{align}
 \langle W_{A_{k_1}, A_{k_2}, \ldots, A_{k_{q}}} \rangle &  =
\exp \left[4 \pi
 \sqrt{\frac{n N}{2(8-N_f)}}N\sum_{a=1}^{q}
 \left(1-\left|1-\frac{k_a}{N}\right|^{3/2}\right)\right] \ ,
\label{antisymmquiver}
\end{align}
where the Wilson loop is in the $k_a$-th anti-symmetric representation for the $a$-th gauge group.
In contrast, for arbitrary configurations of symmetric representations,
we find that the result does not factorize.  We consider
 the special case that
the flavors are distributed uniformly among the gauge groups and the
Wilson loop is in the $k_i$-th symmetric representation for each gauge
group satisfying the constraint that
\begin{equation*}
\frac{k_a}{N}+\frac{9}{4}c_a \qquad \textrm{are independent of}\quad  a
 \ ,
\end{equation*}
where $c_a$ is defined to be $1$ (or $2$)
when the $a$-th gauge group is $USp(2N)$\footnote{By $USp(2N)$ we mean the compact real form of $Sp(2N)$.}.
We then find
\beq
\langle W_{S_{k_1}, \ldots, S_{k_q}}\rangle  =
\exp\left[
  \frac{9\pi }{\sqrt{2(8-N_f)}}n^{\frac{3}{2}} N^{\frac{3}{2}} \left[
\left(1+\frac{4k_{\rm tot}}{9n N}\right)^{\frac{3}{2}}-1
\right]
\right] \ ,
\label{quiversymmresult}
\eeq
where we have introduced $k_{\rm tot} = \sum_{i=1}^q k_i$.  The
qualitative difference between symmetric and anti-symmetric
representations arises in the matrix model from the fact that anti-symmetric representations do not deform the background eigenvalue distribution, while the symmetric representations do.  In the symmetric case, this creates interactions among the eigenvalues and the problem becomes much more involved, except in the case where all of the parameters are distributed symmetrically.

On the gravity side, the representations mentioned
above respectively correspond to
\setlist{nolistsep}
\begin{enumerate}[label=({\arabic*})]
\item fundamental strings
\item D4-branes with $k$ units of electric flux, wrapping AdS$_2$ and an internal $S^3$
\item D4-branes with $k$ units of electric flux, wrapping AdS$_2$ and the space-time $S^3$.
\end{enumerate}
The latter two are the analogues of giant gravitons and dual giant
 gravitons.
Wilson loops in more general representations correspond to multiple such D4-branes.
 For the case $n>1$, the internal space has $q$ independent 3-cycles
 along with their Hodge dual 3-cycles. The general anti-symmetric
 representations labeled by $A_{k_1},A_{k_2},...,A_{k_q}$ correspond to
 $q$ D4-branes, where the $a$-th D4-brane has $k_a$ units of electric
 flux and wraps the $a$-th 3-cycle.  In the symmetric case,
we expect fractional D4-branes, i.e., D6-branes wrapping
space-time $S^3$ cycles and internal blown-up $2$-cycles.
The gravity description of these branes are subtle since
these cycles are of vanishing size,
and possibly requires one to take into account discrete holonomies of
 the B-field on these cycles.
When there are symmetries among the different eigenvalues, as discussed
 above,
then we have a simpler picture,
where there is a single D4-brane wrapping the space-time $S^3$ cycle,
 with $k_{\rm tot}$ units of electric flux. This explains the
formula \eqref{quiversymmresult}.

The paper is organized as follows.  In section \ref{sec.CFT}, we discuss the derivation of
the CFT results.  Section \ref{sec.gravity} contains the dual supergravity description.  We conclude with comments and open problems in
section \ref{sec.discussion}.  We also include appendices on technical material.

\section{Gauge Theory Computations} \label{sec.CFT}

Let us first discuss the supersymmetry preserved by the Wilson loops defined in \eqref{loopdef}.
In the conventions of \cite{Hosomichi:2012ek}, the SUSY variation, which is used for localization, is given by
$\delta A_{\mu} =i \epsilon^{IJ} \xi_I \Gamma_{\mu} \lambda_J, \delta \sigma= -
\epsilon^{IJ} \xi_I \lambda_J$, where $I,J=1,2$ are $SU(2)$ R-symmetry
indices and $\xi_I, \lambda_I$ are $SU(2)$ Majorana spinors.
The SUSY variation of \eqref{loopdef} vanishes if
\begin{align}
\label{WLsusyconst}
\epsilon^{IJ} \xi_I \left( \Gamma_m e_\mu{}^m \dot{x}^{\mu} + \dot{y}
 \right) = 0 \ ,
\end{align}
where $e_{\mu}^m$ is the vielbein.  Multiplying through by $\Gamma_m e_\mu{}^m \dot{x}^{\mu}$ leads to
$|\dot{x}|^2 = \dot{y}^2$.  When the Wilson loop wraps a great circle in $S^5$, \eqref{WLsusyconst} is a projector equation on $\xi_I$ and projects out half of the supersymmetries, with $8$
supersymmetries remaining.\footnote{To see this explicitly, we write the metric on the $S^5$ as
$
ds^2_{S^5} = \left[ \sum_{j=1}^4 \left( \prod_{k=1}^{j-1} \sin^2 \beta_k
 \right) d\beta_{j}^2 \right] + \left( \prod_{k=1}^{4} \sin^2 \beta_k
 \right) d \phi^2 $
 defining $t \equiv \prod_{i=1}^4 \sin \beta_i^0$ and
taking the loop to be a great circle parametrized by $\phi$ with
$\beta_i^0=\mathrm{constant} \, (i=1,\ldots, 4)$, we have $|\dot{x}|^2 = t^2 \Rightarrow \dot y =\pm t$
and \eqref{WLsusyconst} reduces to
$\epsilon^{IJ} \xi_I t \left( \Gamma_5  \pm 1 \right) = 0$ .
We have used the frame
$e_i=\prod_{j<i}\left(\sin\beta_j \right)d\beta_i$ $(i=1,\ldots, 5)$ with $\beta_5\equiv \phi$,
with the other components vanishing.}

In addition to the fermionic supersymmetries, the Wilson loop also preserves the $SU(2)_R \simeq Sp(2)_R$ R-symmetry and
breaks the space-time symmetry to $SO(1,2) \times SO(4)$,
where the $SO(1,2)$ is the conformal group associated with translations
in $\phi$ and the $SO(4)$ is the remaining unbroken rotation group which
leaves invariant the point where the Wilson loop resides in the
transverse space.  These symmetries fit nicely into the supergroup
$D(2,1;2) \times SU(2)$ which has exactly $8$ supersymmetries and is a
subgroup of $F(4)$ (see Table 2.7 in \cite{Dictionary}).  The
specific real forms we are interested in are $F(4;2)$ for Minkowski
signature and $F(4;3)$ for Euclidean signature with subgroup $D(2,1;2;1)
\times SU(2)$ for both cases.\footnote{There is a discrepancy between
Table 3.75 in \cite{Dictionary} and \cite{Parker:1980af}.
The real forms $F(4;2)$ and $F(4;3)$ are listed as having $SL(2,R)$
subgroups in \cite{Dictionary} while in \cite{Parker:1980af}, they
are shown to have $SU(2)$ subgroups.}
Additionally, we note that in the non-orbifold case, the Wilson loops
also preserves an extra $SU(2)_M$ symmetry,
under which the anti-symmetric hypermultiplet transforms as a doublet.
Thus the full symmetry preserved by the half-BPS Wilson loops we consider in this paper is
\begin{align}
D(2,1;2) \times SU(2) \times SU(2)_M \supset SO(1,2) \times
 SO(4)_{\text{space-time}} \times SU(2)_M \times SU(2)_R \ .
\end{align}
The orbifold action will break the $SU(2)_M$ symmetry, however the
Wilson line will remain neutral under this broken symmetry.\footnote{
It is interesting to ask if we could consider a two-parameter
deformation
of the Wilson line which preserves the same supersymmetry,
but charged under $SU(2)_M$
symmetry and $SO(3) \subset SO(4)_{\text{space-time}}$ symmetry, which are not
contained in the $D(2,1;2) \subset F(4)$.}

Let us now move on to the $S^5$ partition function.
The perturbative partition function $Z_{S^5}$ of 5d $\scN=1$ Yang-Mills theory coupled to matter hypermultiplets on the 5-sphere $S^5$ with radius $r$  has been computed in \cite{Kallen:2012va,Kim:2012av} (See also \cite{Kallen:2012cs,Hosomichi:2012ek} for earlier works), building on the localization techniques developed in \cite{Pestun:2007rz,Kapustin:2009kz} for 4d and 3d supersymmetric gauge theories.
By perturbative we mean that the computation does not take into account
the instanton contribution to $Z_{S^5}$.\footnote{See
\cite{Kim:2012av,Kim:2012qf} for the instanton part.}
\smallskip
The result is that the partition function $Z_{S^5}$ reduces to an integration over the Cartan subalgebra of the gauge group, divided by the order of the Weyl group $|\scW|$:
\begin{align}
Z_{S^5} &= \frac{1}{|\scW|} \int_{\textrm{Cartan}} d\sigma \ \Big( ... \Big) \ .
\label{5dZ}
\end{align}
The integrand (the dots in \eqref{5dZ}) is a product of several contributions.
The vector multiplet gives a factor
\begin{align}
e^{-\frac{4 \pi^3 r}{g_{YM}^2} \tr_F(\sigma^2)} \textrm{det}_{\rm Adj}\big(\sinh(\pi \sigma) \, e^{\half f(i \sigma)} \big)
 \ ,
\end{align}
a hypermultiplet in a representation $R$ of the gauge group gives a factor
\begin{align}
\textrm{det}_{R}\big(\cosh(\pi \sigma)^{\frac{1}{4}} \, e^{-\frac{1}{4} f(\half - i \sigma) - \frac{1}{4} f(\half + i \sigma)} \big)
\ ,
\end{align}
a Chern-Simons term with level $k$ contributes a factor
\begin{align}
e^{\frac{\pi k}{3} \tr_F(\sigma^3)} \ \ .
\end{align}
Here $g_{\rm YM}$ is the gauge coupling and $\tr_R$ and $\det_R$ are the
trace and the determinant in the representation $R$. The indices $F$ and
${\rm Adj}$ refer to fundamental and adjoint representations respectively, and the function $f(x)$ is defined by
\begin{align}
f(x) = \frac{i \pi x^3}{3} + x^2\log(1-e^{2i\pi x}) + \frac{i x}{\pi} \Li_2(e^{-2i\pi x}) + \frac{1}{2\pi^2}\Li_3(e^{-2i\pi x}) - \frac{\zeta(3)}{2 \pi^2} \ \ .
\end{align}

\bigskip

We can also incorporate a $\half$-BPS Wilson loop along the
great circle of $S^5$, in the representation $R$ of the gauge group.
In the localization computation of the
partition function \cite{Kallen:2012va,Kim:2012av}, the saddle
point equations imply $A=0$ and $\sigma$ constant,\footnote{There are
other saddle points with non-trivial profile of gauge fields, however
these correspond to instanton contributions which does not change
the leading large $N$ analysis in this paper.
}
and hence the Wilson loop operator \eqref{loopdef}
reduces to an insertion of the following exponential factor to the integrand of the matrix integral:
\begin{align}
\tr_R \big( e^{2\pi \sigma} \big) \ \ .
\end{align}

The $S^5$ partition function depends on the value of the gauge coupling
constant $g_{\rm YM}$, which induces a relevant deformation
of the UV fixed point.
To discuss the UV fixed point we consider the limit
where such a deformation is completely turned off:
\begin{align}
g_{\rm YM}^2 \gg r \ .
\end{align}
Moreover for the comparison with gravity we take the large $N$
limit
\begin{align}
N \gg 1 \ ,
\end{align}
where $N$ is the dimension of the Cartan subalgebra (number of integration variables).
In these limits the contributions from instantons and the contribution
from the Yang-Mills kinetic term $e^{-\frac{4 \pi^3 r}{g_{\rm YM}^2}
\tr_F(\sigma^2)}$ are subleading \cite{Jafferis:2012iv},
and hence will be neglected
 in the rest of the computations.\footnote{Note that we are not taking
 the 't Hooft limit; there will be no dependence on $g_{\rm YM}$ for
 the rest of the paper and we concentrate on the $N$ dependence.}

After taking into account these considerations, we simplify the matrix integral as
\begin{align}
Z_{S^5} &= \frac{1}{|\scW|} \int_{\textrm{Cartan}} d\sigma \, \, e^{-F(\sigma)} \ ,
\end{align}
where in the large $|\sigma|$ limit we have
\begin{align}
F(\sigma) &= \tr_{\rm Adj} F_V(\sigma) + \sum_j \tr_{R_j} F_H(\sigma) \ ,
\end{align}
with
\begin{align}
F_V(\sigma) = \frac{\pi}{6}|\sigma|^3 - \pi|\sigma|  \ , \quad
F_H(\sigma) = -\frac{\pi}{6}|\sigma|^3 - \frac{\pi}{8}|\sigma| \ .
\label{Fapprox}
\end{align}

\subsection{Fundamental Representation}
Let us first consider the theory with $n=1$.
This theory is 5d $\scN=1$ $USp(2N)$ gauge theory with $N_f$
hypermultiplets in the fundamental representation and one hypermultiplet
in the antisymmetric representation.  In this case the matrix integral
is over $N$ real parameters
$\sigma_j$ ($j=1,...,N$), parametrizing the Cartan as
$\{ \sigma_1, \ldots, \sigma_N, -\sigma_1, -\sigma_N\}$.
We will evaluate the matrix integral in the saddle point approximation,
where, as we will justify later, the saddle point value of the eigenvalues of order $\ord(N^{1/2})$.
This means that in our large $N$
approximation we can take the large $|\sigma_j|$
limit inside the function $F(\sigma)$,
and we have
\begin{align}
F(\sigma) &= \sum_{i\neq j} \Big( F_V(\sigma_i - \sigma_j) + F_V(\sigma_i + \sigma_j) + F_H(\sigma_i - \sigma_j) + F_H(\sigma_i + \sigma_j) \Big)  \no\\
  & + \sum_j \Big( F_V(2\sigma_j) + F_V(-2\sigma_j) + N_f \, F_H(\sigma_j) + N_f \, F_H(-\sigma_j) \Big) \ .
\end{align}
The Weyl group of $USp(2N)$ is given by $\scW=\mathfrak{S}_N\ltimes
\bZ_2^N$,
and hence
$|\scW |=N! 2^N$.

The large $N$ limit of the free energy $F_{S^5} = - \log\left|Z_{S^5}\right|$ in the saddle point approximation of matrix models
is given in \cite{Jafferis:2012iv}:
\begin{align}
F_{S^5} = -\frac{9 \sqrt{2} \pi}{5\sqrt{8-N_f}} \, N^{5/2} +{\cal O}(N^{5/2}) \
 .
\label{FCFT}
\end{align}
We will comment on the holographic computation of this formula in section \ref{sec:Fgravity}.

Here we study the fundamental $\half$-BPS Wilson loop $W_{\rm fund}$,
whose VEV in the
limit $\frac{r}{g_{YM}^2} \ll1$ is given by
\begin{align}
 \langle W_{\rm fund} \rangle &= \frac{1}{Z_{S^5}} \frac{1}{|\scW|}\int d^N \sigma \,
 \frac{1}{2N}\left[\sum_{j=1}^{N} (e^{2\pi \sigma_j} + e^{-2\pi
 \sigma_j}) \right] \, e^{-F(\sigma)} \nonumber \\
&= \frac{1}{Z_{S^5}} \frac{1}{|\scW|}\int d^N \sigma \,
 \frac{1}{N}\left[\sum_{j=1}^{N} e^{2\pi \sigma_j} \right] \, e^{-F(\sigma)}  \ .
\end{align}
We are looking for the saddle point of this integral in the large $N$
limit. We will assume as in \cite{Jafferis:2012iv} that the saddle point
is given by $\sigma_j^{\star} = N^{\alpha} x_j$ with the saddle point
variables $x_j$ of order $\ord(N^0)$. We also assume that the variables
$x_j$ at the saddle point condense into a continuous distribution, $\rho(x)$, which is
smooth on an interval of finite length $L$
and zero outside the interval.
These assumptions will be justified
in the computation that follows.

We can then replace the $N$ variables by a continuous variable $x$ with density $\rho(x)$:
\begin{align}
 \rho(x) = \frac{1}{N} \sum_{j=1}^N \delta(x- x_i) \ ,\quad \ \int dx\,
 \rho(x) = 1 \ .
\label{rhonormal}
\end{align}
In this limit the Wilson loop becomes
\begin{align}
\label{Wilsoncontinuum}
\langle W_{\rm fund} \rangle &= \frac{1}{Z_{S^5}} \frac{1}{|\scW|}\int
 D\rho \, \left[ \int \! dx \, \rho(x)\, e^{2\pi N^{\alpha} x}
 \right] \, e^{-F[\rho,\mu]} \no \ , \\
 F[\rho,\mu] &= -\frac{9\pi}{8}N^{2+\alpha} \int dx \,  dy \, \rho(x)\rho(y)(|x-y|+|x+y|) + \frac{\pi(8-N_f)}{3}N^{1+3\alpha} \int dx \, \rho(x) |x|^3 \no\\
 & + \mu \Big(1- \int dx \, \rho(x) \Big) \ .
\end{align}
where we have added a Lagrange multiplier $\mu$ to impose the constraint $\int dx \, \rho(x) =1$.
We have $\int dx\, \rho(x)e^{2\pi
N^{\alpha} x} \sim \ord(e^{2\pi N^{\alpha} L}) $,
which in the saddle point approximation is subleading compared with
other contributions in $F[\rho, \mu]$.  The Wilson loop therefore does not affect the saddle point equations.

The saddle point equation reduces to
\begin{align}
\label{saddleqn}
 0 = \frac{\delta F[\rho]}{\delta \rho(x)} = -\frac{9\pi}{4}
 N^{2+\alpha} \int  dy \, \rho(y)(|x-y|+|x+y|) +
 \frac{\pi(8-N_f)}{3}N^{1+3\alpha} |x|^3 - \mu \ .
\end{align}
Non-trivial solutions are obtained when the two terms are of the same
 order, namely when $\alpha = \half$; only in this case the mutual
repulsion among the eigenvalues balances the attraction from the cubic
 potential, giving continuous eigenvalue distributions as assumed
 previously.
In this case the two first terms in $F[\rho, \mu]$ are both of order
 $N^{5/2}$,
which justifies a posteriori the assumption that the Wilson loop factor
does not affect the saddle point equation.

It is easy to realize that $F[\rho]$ only depends on the even part of $\rho$ ($=\half[\rho(x)+\rho(-x)]$). So the integration over $\rho$ can be reduced to the integration over even $\rho$ up to a factor coming out of the integration measure, which does not affect the Wilson loop computation (because of the normalization of the Wilson loop).
Differentiating twice the equation \eqref{saddleqn} with respect to $x$,
 and assuming an even distribution $\rho$, we get
\begin{align}
 \rho(x) = \frac{|x|}{x_0^2} \quad \textrm{for} \quad -x_0 < x < x_0 \ ,
 \quad x_0^2 := \frac{9}{2(8-N_f)} \ ,
\label{rhosad}
\end{align}
which satisfies the normalization condition \eqref{rhonormal}.
Plugging this back into the expression \eqref{Wilsoncontinuum} we get at leading order in $N$:
\begin{align}
\label{Wfund}
 \langle W_{\rm fund} \rangle &=  \frac{\exp(2\pi x_0 N^{1/2})}{2 \pi x_0 N^{1/2}}\no\\
 & = \frac{\sqrt{2(8-N_f)}}{6 \pi N^{1/2}} \, \exp \left(\frac{6\pi
 N^{1/2}}{\sqrt{2(8-N_f)}} \right) \ ,
\end{align}
which is the advertised result \eqref{agreementf}.
It is rather simple to understand the leading contribution: the Wilson
loop contribution $e^{2\pi \sigma}$ is maximized when the eigenvalues
$\sigma$ take the maximal possible value, which is $N^{1/2} x_0$.

\subsection{Anti-symmetric Representations}\label{CFT.antisymm}

Let us next consider the $k$-th anti-symmetric representation $A_k$ of
the $USp(2N)$ gauge group.
The Wilson loop in representation $A_k$ is given in the matrix model by
\begin{align}
 \langle W_{A_k} \rangle &=  \frac{1}{Z_{S^5}} \frac{1}{|\scW|} \, \binom{2N}{k}^{-1} \int d^N \sigma \sum_{1\le j_1< j_2<\ldots < j_k \le 2N}
e^{2\pi(\sigma'_{j_1}+\sigma'_{j_2}+\ldots +\sigma'_{j_k})} \,
 e^{-F(\sigma)}  \ ,  \no
\end{align}
where $\sigma'_{j} = \sigma_j$ and $\sigma'_{N+j} = - \sigma_j$ for
$j=1,\ldots,N$.

In the sum $ \sigma'_{j_1}+\sigma'_{j_2}+\ldots +\sigma'_{j_k} $ it is
possible that some terms cancel with each other.
In particular this expression is invariant under exchanging $k$ with $2N-k$, so that $ \langle W_{A_k} \rangle = \langle W_{A_{2N-k}} \rangle $. Hence we need only consider $1 \le k \le N$.

Let us first consider the $k$-plets $(j_1,...,j_k)$ (with $k \le N$) such that there is no cancellation in $ \sigma'_{j_1}+\sigma'_{j_2}+\ldots +\sigma'_{j_k} $ (which means all $\sigma_j$ are different). These terms contribute a factor $ \binom{2N}{k}^{-1} I_k $ with
\begin{align}
 I_k &=  \frac{1}{Z_{S^5}} \frac{1}{|\scW|} \int d^N \sigma \, \sum_{\epsilon_1,\ldots,\epsilon_k=\pm 1}\sum_{1\le j_1<\ldots < j_k \le N}
e^{2\pi(\epsilon_1\sigma_{j_1}+ \epsilon_2\sigma_{j_2}+ \ldots +\epsilon_k\sigma_{j_k})} \, e^{-F(\sigma)} \ . \no
\end{align}
The symmetry of $F$ implies that all terms in the sum over $\epsilon_1,\ldots,\epsilon_k $ produce the same contribution, so that
\begin{align}
\label{I_k}
 I_k &=  \frac{1}{Z_{S^5}} \frac{1}{|\scW|} 2^k \binom{N}{k} \int d^N \sigma \, e^{2\pi(\sigma_1+ \sigma_2+ \ldots +\sigma_k)}  \, e^{-F(\sigma)}
\end{align}
Again we assume that the saddle point eigenvalues are distributed along an interval of length of order $N^{\alpha}$ with $\sigma^{\star}_j = N^{\alpha} x_j$.

We can again argue that the Wilson loop does not modify the saddle
point \eqref{rhosad} in the large $N$ limit;
the Wilson loop operator contributes at most a term of order
$N^{1+\alpha}$ to the saddle point equation and this is again subleading
compared to the term coming from $F(N^{\alpha} x_j)$ with
$\alpha=\half$.
However, this does not mean that the answer is $k$ times the
fundamental representation. This is because we need to choose
$k$ distinct eigenvalues $\sigma_1, \ldots, \sigma_k$ from the
eigenvalue distribution, and therefore we cannot
always take the maximal value $\sigma_j=N^{1/2} x_0$ when $k$ is large.

\begin{figure}[htbp]
\centering{\includegraphics[scale=0.6,trim=0 140 0 150]{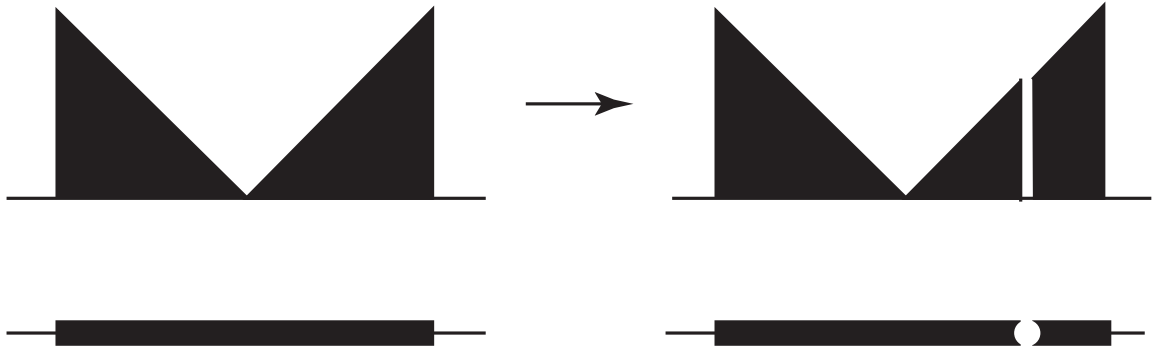}}
\caption{Insertion of a Wilson line in anti-symmetric representation
shifts part of the eigenvalues by a constant, or equivalently an
excitation of a ``hole'' in the eigenvalues.}
\label{fig.antisymm}
\end{figure}

The dominant contribution to the integral \eqref{I_k} comes from
configurations when the first $k$ eigenvalues cover an interval
$[x_0 \cos\theta_k, x_0]$ at the right end of the saddle point distribution
$\rho$ so that the factor $e^{2\pi(\sigma_1+ \sigma_2+ \ldots
+\sigma_k)}$ attains maximum.
Here the angle $\theta_k \in [0,\pi/2]$ is chosen such that we indeed have
$k$ eigenvalues in the interval:
\beq
k = 2N \int_{x_0 \cos\theta_k}^{ x_0} \rho(x) \, dx = N \sin^2 \theta_k \ .
\label{ktheta}
\eeq
Intuitively, the Wilson line operator
 corresponds to a constant electric flux for
$k$ of the eigenvalues, hence shifting the $k$ eigenvalues
 and creating a ``hole'' in the eigenvalue
distribution (cf. \cite{Yamaguchi:2006tq}), see Figure
\ref{fig.antisymm}.

The maximal value for the $i$-th eigenvalue $\sigma_i$
is $N^{1/2} x_0 \cos\theta_i$, and hence contributes $e^{2\pi N^{1/2}
x_0 \cos \theta_i}$ to the integral.
We then evaluate \eqref{I_k} by summing over these contributions:
\begin{align}
I_k=2^k \binom{N}{k}\exp\left(\sum_{i=1}^k\, 2\pi x_0 \cos\theta_i
N^{1/2} \right)\no
\simeq 2^k \binom{N}{k}\exp\left(\int_{0}^{\theta_k}\! \! d\theta_k \frac{\partial k}{\partial \theta_k} 2\pi x_0 \cos\theta_i
N^{1/2} \right) \ .
\end{align}
This gives
\begin{align}
 I_k &= \ 2^k \binom{N}{k}  \, \exp\left[\frac{4\pi}{3} x_0
 N^{3/2}(1-\cos^3 \theta_k)\right] \no \\
&= 2^k \binom{N}{k}  \, \exp\left[\frac{4\pi}{3} x_0 N^{3/2} \left( 1-\left(1-\frac{k}{N}\right)^{3/2} \right)\right] \ ,
\end{align}
where the prefactor $2^k \binom{N}{k}$
gives only a subleading correction of order $N$ to the exponent.

Now we consider the terms in the sum over $j_1,...j_k$ such that two
$\sigma'_j$ cancel. These terms will contribute a factor
$\binom{2N}{k}^{-1} (N-k+2)I_{k-2}$. From the previous explanation it
follows that this contribution is suppressed, as compared to the
contribution $\binom{2N}{k}^{-1} I_k$, by a factor of
order $e^{-2 (2\pi N^{1/2} x_0 \cos \theta_k)}$.\footnote{When $k \rightarrow N$ we have $\cos\theta_k \rightarrow 0$ and the contribution is not suppressed, however it leads to the same contribution as $I_k$ and the sum over all the contributions reduce to the same leading term in the exponent.}
Similarly, all the other terms left in the sum over $j_1,...,j_k$ are also subleading. Thus the leading contribution in the large $N$ limit is $\binom{2N}{k}^{-1} I_k $. Explicitly we have
\begin{align}
\log \, \langle W_{A_k} \rangle = \frac{4\pi}{3} x_0 N^{3/2} \left[1-\left| 1-\frac{k}{N} \right|^{3/2} \right] \ ,
\end{align}
which coincides with \eqref{agreementa}.\footnote{The subleading correction is of order $N$ when $k$ and $N$ are of the
same order.}
This expression is valid for $1 \le k \le 2N$, ensuring $\langle W_{A_k} \rangle$ = $\langle W_{A_{2N-k}} \rangle$.
As a consistency check, when $k \ll N$ we have
\beq
\log \, \langle W_{A_k} \rangle \sim k(2 \pi N^{1/2} x_0) \ ,
\label{Akresult}
\eeq
which could be interpreted as the $k$ times the fundamental string
contribution \eqref{Wfund}. Of course, this follows directly
from the derivation presented above.

As explained in Appendix \ref{sec.grouptheory}, the
anti-symmetric representation $A_k$ defined in introduction is
a reducible
representation, and in particular (when $1\le k\le N$) contains the irreducible
representation
defined by the Young diagram with a single column with $k$ boxes.
The arguments similar to those already explained in this subsection
shows that the contributions from other irreducible
representations are exponentially
suppressed, and the leading contribution comes from this irreducible representation.

\subsection{Symmetric Representations}\label{CFT.symm}

Let us move onto the case of $k$-th symmetric representation $S_k$ of $USp(2N)$.
We have
\begin{align}
 \langle W_{S_k} \rangle &=  \frac{1}{Z_{S^5}} \frac{1}{|\scW|} \, \binom{2N+k-1}{k}^{-1}
 \int d^N \sigma \sum_{1\le j_1\le  j_2 \le\ldots \le j_k \le 2N}
e^{2\pi(\sigma'_{j_1}+\sigma'_{j_2}+\ldots +\sigma'_{j_k})} \,
 e^{-F(\sigma)}   \ .
\end{align}
In the sum $\sigma'_{j_1}+\sigma'_{j_2}+\ldots +\sigma'_{j_k}$
some of the terms could cancel out from the expression, however
these give only exponentially suppressed contributions,
by the reason already explained in the case of anti-symmetric representations.
If we neglect these contributions we have
\begin{align*}
\frac{1}{Z_{S^5}} \frac{1}{|\scW|} \, \binom{2N+k-1}{k}^{-1}2^k
 \int d^N \sigma \sum_{1\le j_1\le  j_2 \le\ldots \le j_k \le N}
e^{2\pi(\sigma_{j_1}+\sigma_{j_2}+\ldots +\sigma_{j_k})} \,
 e^{-F(\sigma)}   \ .
\end{align*}
The summation here still contains several different terms.
If we denote the partition of $k$ by $\mu=(\mu_1, \ldots, \mu_l), \sum \mu_i=k$,\footnote{
Readers should not confuse this partition with a partition specifying
a representation of $USp(2N)$. Rather it actually corresponds to a
symplectic semi-standard Young tableaux in the language of Appendix \ref{sec.grouptheory}.
Here we have avoided use of such terminologies for the minimality of the explanation.
}

then
\beq
 \sum_{1\le j_1\le  j_2 \le\ldots \le j_k \le 2N}
e^{2\pi(\sigma_{j_1}+\sigma_{j_2}+\ldots +\sigma_{j_k})}
=\sum_{\mu}\left( e^{2\pi \sum_i \mu_i \sigma_i} + {\rm cyclic} \right)\ .
\eeq
This contains many different contributions. For example, the
contribution from the partition $\mu=(1,
\ldots, 1)$ is the same as that from the anti-symmetric representation \eqref{Akresult},
except for the overall factor of $\binom{2N+k-1}{k}^{-1}$:
\beq
\langle W_{S_k} \rangle\Big|_{\mu=(1, \ldots,
1)}=\binom{2N+k-1}{k}^{-1}\binom{2N}{k} \,
\langle W_{A_k} \rangle \ .
\eeq

This is not the only contribution, however.
On the other extreme there is a contribution from $\mu=(k)$, i.e.
\begin{align}
\langle W_{S_k} \rangle\Big|_{\mu=(k)}=
\frac{1}{Z_{S^5}} \frac{1}{|\scW|} \,
 \binom{2N+k-1}{k}^{-1} 2^k N \int d^N \sigma \,
e^{2\pi k \sigma_1} \,
 e^{-F(\sigma)}   \ .
\end{align}
This is the contribution from the
``large winding Wilson loop''.

\begin{figure}[htbp]
\centering{\includegraphics[scale=0.6,trim=0 160 0 150]{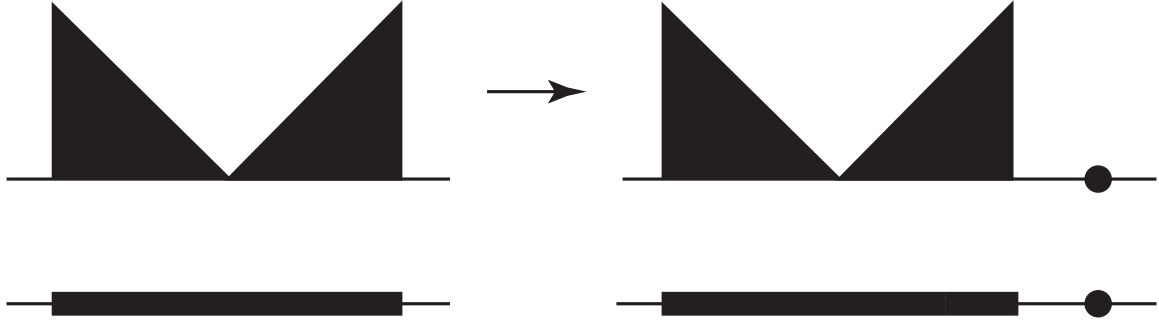}}
\caption{Insertion of a Wilson line in symmetric representation
corresponds to exciting one of the eigenvalues to a large value,
or equivalently an
 excitation of a ``particle'' in the Fermi sea.}
\label{fig.symm}
\end{figure}

Let us evaluate this contribution.
We can replace all the $\sigma_i$ by the continuum distribution
determined by \eqref{rhosad}, except for $\sigma_1$.
Since there is a the factor of $k$ multiplying $\sigma_1$
and since $k$ can be large, the Wilson line does affect the saddle
point for $\sigma_1$.
This happens when $k$ is of order $N$ or larger;
the leading contribution of free energy is of order $N^{5/2}$,
however they cancel out when we compute the Wilson loops (due to the
normalization factor $Z_{S^5}$), and the subleading contribution of
order
$N^{3/2}$ becomes comparable with the Wilson loop contribution of order
$kN^{1/2}$, when $k$ is of order $N$.
Intuitively, the eigenvalue $\sigma_1$ moves inside the effective potential
created by the other background eigenvalues,
and can be regarded as a ``particle'' in the eigenvalue distribution
(Figure \ref{fig.symm}).

We now have
\beq
  \binom{2N+k-1}{k}^{-1} 2^k N \frac
{\displaystyle\int d\sigma_1 \exp\left[-F_{\rm eff}(\sigma_1;k)\right]}
{\displaystyle\int d\sigma_1 \exp\left[-F_{\rm eff}(\sigma_1;k=0)\right]}
\ ,
\label{sigma1int}
\eeq
with
\begin{align}
&F_{\rm eff}(\sigma_1;k)=-2\pi k \sigma_1
+2 \left[F_V(2\sigma_1)+N_f F_H(\sigma_1) \right] \\
&\quad +2 N \int \!dy\,
\rho(y) \left[ F_V(\sigma_1-N^{\half}y)+F_V(\sigma_1+N^{\half}y)
+F_H(\sigma_1-N^{\half}y)+F_V(\sigma_1+N^{\half}y)\right] \ , \no
\end{align}
where the factors $2$ comes is due to the property $F_{V,H}(-\sigma)=F_{V,H}(\sigma)$.
We will evaluate the integral \eqref{sigma1int}
by the saddle point approximation with respect to $x_1 := x_0^{-1}N^{-1/2}\sigma_1$.
We assume that the saddle point is given by $x_{1, *} > 1$. This will be justified a posteriori by the result of our computation. Under this assumption we have
\beq
F_{\rm eff}(x_1;k) \simeq \pi x_0 N^{3/2} \left[ - x_1 \left(2 \frac{k}{N}
+\frac{9}{2} \right) + \frac{3}{2} x_1^3  \right] \ ,
\label{Feff}
\eeq
where we used
\begin{equation}
\begin{split}
\int dy \, \rho(y) \left[F_V(\sigma_1-N^{\half} y)+F_V(\sigma_1+N^{\half}y)\right]
& =2\left[\frac{\pi}{6} (N^{\half}x_0)^3 \left(x_1^3+\frac{3}{2}x_1\right)
-\pi (N^{\half}x_0) x_1
\right] \ , \\
\int dy \, \rho(y) \left[F_H(\sigma_1-N^{\half} y)+F_H(\sigma_1+N^{\half}y)\right]
& =2\left[-\frac{\pi}{6} (N^{\half}x_0)^3 \left(x_1^3+\frac{3}{2}x_1\right)
-\frac{\pi}{8} (N^{\half}x_0) x_1
\right] \ .
\label{FVHint}
\end{split}
\end{equation}
This is extremized by
\beq
x_{1, *} = \sqrt{\frac{4k}{9N}+1} \ ,
\eeq
which justifies our previous assumption.
Note that the eigenvalue $\sigma_{1, *} = N^{1/2}x_0 x_{1, *}$ is outside the range
occupied by other eigenvalues (Figure \ref{fig.symm}).

At the saddle point we have the contribution to the free energy
\begin{align}
\log \langle W_{S_k} \rangle\Big|_{\mu=(k)} &=-\left(F_{\rm eff}(\sigma_{1,*};k)-
F_{\rm eff}(\sigma_{1,*};k=0)\right) \nonumber \\
&=\frac{9\pi }{\sqrt{2(8-N_f)}}N^{3/2}
\left[
\left(1+\frac{4k}{9N}\right)^{3/2} -1\right]\ .
\end{align}

In general there are many contributions from various different choices
of $\mu$, and we need to take all of them into account.
For example, when $k$ is small all of them has the same leading
contribution, with $\mu=(1, \ldots, 1)$ having the largest subleading
correction due to the largest multiplicity $\binom{N}{k}$.
However when $k$ is large, of order $N$ or larger,
we can verify from the expressions above that
the contribution from $\mu=(k)$ dominates.

Some readers might worry about contributions
from other $\mu$, say $\mu=(k-1, 1)$.
However when $k$ is large this is suppressed by an exponential factor
(cf. \cite{Yamaguchi:2007ps}):\footnote{The presence of this factor
ensures that our Wilson loop is not simply the multiple wound string,
but a loop in the symmetric representation.}
\beq
\exp\left[ -\frac{\partial}{\partial k} \frac{9\pi }{\sqrt{2(8-N_f)}}N^{\frac{3}{2}}
\left[
\left(1+\frac{4k}{9N}\right)^{\frac{3}{2}} -1\right] \right]
=
\exp\left[-
\frac{6\pi}{\sqrt{2(8-N_f)}} N^{\frac{1}{2}}
\sqrt{1+\frac{4k}{9N}}
\right] \ .
\eeq
Similarly, we can check that contributions from other $\mu$ are likewise
exponentially suppressed.

Therefore we finally have
\begin{align}
\log \langle W_{S_k}\rangle=\log \langle W_{S_k} \rangle\Big|_{\mu=(k)}
=\frac{9\pi }{\sqrt{2(8-N_f)}}N^{3/2}
\left[
\left(1+\frac{4k}{9N}\right)^{3/2} -1\right]\ ,
\end{align}
as claimed in \eqref{agreements}.
Let us note that
\beq
\frac{\log \langle W_{S_k} \rangle}{\log\langle W_{\rm fund} \rangle}
= \frac{3N}{2} \left[
   \left(1+\frac{4k}{9N}\right)^{3/2}-1
\right] \ .
\label{symresult}
\eeq
For $k\ll N$, this reduces to
\beq
\frac{\log \langle W_{S_k} \rangle}{\log \langle W_{\rm fund} \rangle}
=k \ ,
\eeq
as expected.

\subsection{General Representations}\label{CFT.general}

We can consider more general representations (see Appendix
\ref{sec.grouptheory} for representation theory of $USp(2N)$).
An irreducible representation of $USp(2N)$ can
be labeled by a Young diagram with at most
$N$ rows.
We can label the representation by a partition $k=(k_1, \ldots,
k_m)$, or its dual (transpose) by $k^T=:l=(l_1, \ldots, l_{k_1})$.

The Wilson line operator in representation $R$
corresponds to an insertion of
\beq
sp_{(k_1, \ldots, k_m)}(e^{2\pi \sigma_1}, \ldots, e^{2\pi \sigma_{N}}) \ ,
\eeq
where $sp_k$ is the symplectic character
for representation, introduced in Appendix \ref{sec.grouptheory}.

\begin{figure}[htbp]
\centering{\includegraphics[scale=0.6,trim=0 140 0 150]{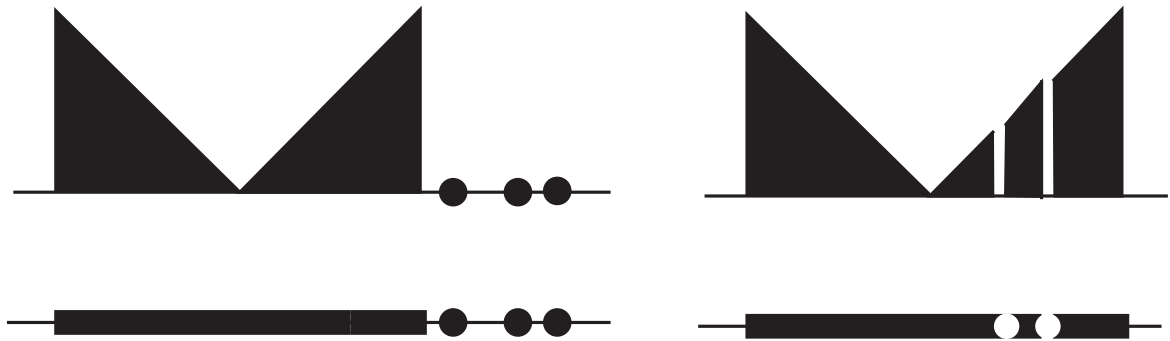}}
\caption{Wilson lines in general representations can be described
either as the excitation of several non-interacting particles above the Fermi sea (left)
or the excitation of non-interacting holes (right). The two descriptions
correspond to a Young diagram and its dual, i.e. taking the transpose could be thought as a
Bogoliubov-like transformation.}
\label{fig.genrep}
\end{figure}

Let us first consider the representation described by
$k=(k_1, \ldots, k_m)$. We assume that $m$ is a finite number
which stays constant when $N$ grows large.
However there are no restrictions on the size of $k_a$'s, and they can
grow with some power of $N$.
The symplectic character $sp_{\lambda}(x)$ is a sum of
terms labeled by symplectic semistandard Young tableaux \eqref{spassum}.
By the argument similar to the previous subsections, we can argue that
the leading contribution comes from the tableaux
\begin{equation*}
{\rm \young(11111\ybullet\ybullet
 1,2222\ybullet\ybullet 2,33\ybullet\ybullet3,\ybullet\ybullet,\ybullet\ybullet,mm)} \ .
\end{equation*}
This corresponds to an insertion of
\begin{equation*}
e^{2\pi \sum_a k_a \sigma_a}
\end{equation*}
into the integrand of the matrix model.
Again, there are multiplicities associated to this factor which do not
alter the leading contribution and hence will be neglected for the rest
of the computation.
When all the $k_a$ are large (of order $N$ or larger),
this factor excites $N$ eigenvalues  $\sigma_1, \ldots, \sigma_m$ out of the
Fermi sea and
we can then write down the effective matrix model
for $\sigma_1, \ldots, \sigma_m$.
The integrand of this matrix model is
$e^{-F_{\rm eff}(\sigma_a)}$ with
\begin{equation}
\begin{split}
F_{\rm eff}(\sigma_a) =-2\pi& \sum_{a} k_a \sigma_a+\sum_{a}
 2 \left( F_V(2 \sigma_a)+ N_f F_H(\sigma_a) \right) \\
&+\sum_{a<b}\left(F_V(\sigma_a-\sigma_b)+F_V(\sigma_a+\sigma_b)
+F_H(\sigma_a-\sigma_b)+F_H(\sigma_a+\sigma_b)\right) \\
&+
2 \sum_{a}\int \! dy \, \rho(y)\Big[
F_V(\sigma_a-N^{\half}y)+F_V(\sigma_a+N^{\half}y) \\
& \hspace{4cm} +F_H(\sigma_a-N^{\half}y)+F_H(\sigma_a+N^{\half}y)
\Big] \ .
\end{split}
\end{equation}
This evaluates to (recall \eqref{FVHint})
\beq
F_{\rm eff}(x_a)
\simeq \pi x_0 N^{3/2} \sum_{a}
\left[
-x_a\left(2\frac{k_a}{N}+\frac{9}{2}\right)
+\frac{3}{2} x_a^3
\right] \ ,
\eeq
where we neglected the subleading corrections and
we defined $x_a:=x_0^{-1} N^{-\half}\sigma_a$.
Note that interactions among $\sigma_a$'s are subleading of order
$N^{\half}$, hence at this order the eigenvalues behave independently;
the excitations from the Fermi sea behave as non-interacting particles
in the leading order (Figure \ref{fig.genrep}).

We can extremize $F_{\rm eff}(x_a)$ to obtain
\beq
x_{a,*}=\sqrt{\frac{4k_a}{9N}+1} \ ,
\eeq
and the Wilson line evaluates to
\beq
\log\, \langle W_{(k_1, \ldots, k_m)}\rangle=
\frac{9\pi }{\sqrt{2(8-N_f)}}N^{3/2} \sum_{a=1}^m
 \left[
\left(1+\frac{4 k_a}{9N} \right)^{3/2}-1
\right] \ .
\eeq

\bigskip
We can also consider a similar situation, where
this time the transpose of the Young diagram,
$l:=k^T$ takes the form
$l=(l_1, \ldots, l_m)$ with $m$ a finite number.
Note that we have $l_a\le N$, however $l_a$
can still be of order $N$.

In the symplectic character $sp_{\lambda}(x)$,
the leading contribution in this case comes from the
tableaux
\begin{equation*}
{\rm \young(111\ybullet\ybullet1,222\ybullet\ybullet \lm,33\ybullet,4\ybullet\ybullet,5\ybullet\lthree,\ybullet\ltwo,\ybullet,\lone)} \ .
\end{equation*}
From the $a$-th column, there is a contribution of the form
\begin{equation*}
\sum_{j_1<j_2< \ldots <j_{l_a}} e^{2\pi
 (\sigma_{j_1}+\sigma_{j_2}+\ldots +\sigma_{j_{l_a}})}\ ,
\end{equation*}
which contributes a leading contribution which coincides
with that from the $l_a$-th anti-symmetric representation
$A_{l_a}$.
Summing up these contributions over $a$,
we find
\beq
\log\, \langle W_{(l_1, \ldots, l_m)^T}\rangle=
\frac{4\pi }{2\sqrt{8-N_f}}N^{3/2} \sum_{a=1}^m
 \left[
\left(1-\frac{ k_a}{N} \right)^{2/3}-1
\right] \ .
\eeq
This again takes a factorized form,
and has an interpretation as excitation of non-interacting holes inside
the Fermi sea (Figure \ref{fig.genrep}).
Note that factorization breaks down in the subleading order since
the multiplicity of the leading contribution, which contributes to the
subleading correction, is affected by the presence of the neighboring columns.

For more general representations,
we can understand Wilson lines in two descriptions,
those as a generalization of symmetric representations or anti-symmetric
representations.
However, strictly speaking neither description is completely justified
when we have a Young diagram $k=(k_1, \ldots, k_m)$
and both $m$ and $k_1$ grow
with some  power of $N$.

On the gravity dual discussed in the next section,
the two descriptions, particles or holes, correspond to\footnote{The related discussion for circular Wilson loops for 4d $\scN=4$
theory can be found in \cite{Gomis:2006sb}, where
combinatorial formulas, Giambelli's formula and Jacobi-Trudi
formula, played crucial roles. Analogous formulas
are known for symplectic groups.}

\begin{enumerate}[label=(\arabic*)]
\item multiple D4-branes wrapping $AdS_2$ and spacetime $S^3$,
\item multiple D4-branes wrapping $AdS_2$ and internal $S^3$.
\end{enumerate}

\subsection{Quiver Theories} \label{CFT.quiver}

In \cite{Jafferis:2012iv} more general quiver-type gauge theories were
considered depending on an integer $n$. The gauge group is
$G = USp(2N) \times SU(2N)^p$ for $n=2p+1$ and $G = USp(2N) \times
SU(2N)^{p-1} \times USp(2N) $ or $SU(2N)^p$  for $n=2p$. The matter
content is given by a bifundamental hypermultiplet in each pair of
adjacent gauge groups, one antisymmetric hypermultiplet in each external
$SU(2N)$ (gauge factor at the beginning or the end of the line quiver
picture) and $N_f^a$ fundamental hypermultiplets in the $a$-th gauge
group factor, with $N_f = \sum_a N_f^a$.

The saddle point of the corresponding matrix model has been analyzed in
\cite{Jafferis:2012iv}. They assume that all the eigenvalues scale as
$N^{\alpha}$,
just as in $n=1$. The integrand of the matrix model contain terms of
order $N^{2+3\alpha}$, which is extremized by the ansatz
\begin{equation}
\begin{split}
&\sigma'^{(a)}_i=\sigma'_i \  \quad (a=1, \ldots, p) \ , \\
&\sigma'_i=-\sigma'_{N+i} \quad (i=1, \ldots, N, \quad a=1,
\ldots, p) \ .
\label{quiversaddle}
\end{split}
\end{equation}
However the extremal value of $N^{2+3\alpha}$ term vanishes under
\eqref{quiversaddle}, and we need to discuss subleading terms,
which are given by
\begin{align}
 F[\rho] &= -\frac{9\pi}{8} \,n \, N^{2+\alpha} \int dx \,  dy \,
 \rho(x)\rho(y)(|x-y|+|x+y|) + \frac{\pi(8-N_f)}{3}N^{1+3 \alpha} \int
 dx \, \rho(x) |x|^3 \ ,
\label{eq123}
\end{align}
which is identical to the leading free energy for the
$USp(2N)$ theory, up to a factor of $n$.
This means that we again have $\alpha=1/2$, and that at the saddle point the matrix model is the same as the
matrix model of $USp(2N)^n$ gauge group without bifundamentals
\cite{Jafferis:2012iv}.\footnote{In the discussion above we have assumed that $n$ is small. However it is
also possible to take $n$ large, for example $n=n' N^{\beta}$ with $n'$ finite.
In this case the leading contribution is of order $n N^{7/2}$, which vanish
under \eqref{quiversaddle}. However in the next order \eqref{eq123}
gives $\alpha=(1+\beta)/2$, and hence the free energy scales as $O(n N^{5/2})=O(N^{(5+3\beta)/2})$.}

\bigskip

We can compute the VEV of Wilson loops in these
theories.
For the $a$-th gauge group (either $USp(2N)$ or $U(2N)$)
we could turn on a Wilson line in
representation $R_a$, and compute its expectation values
$\langle W_{R_1, R_2, \ldots, R_{q}}\rangle$. Here we take $R_a$
to be either a anti-symmetric representation $A_{k_a}$ or a
symmetric representation $S_{k_a}$, and $q$ is the total number of
gauge groups, i.e., $q=p+1$ for $n=2p+1$ and $q=p$ or $q=p+1$ for $n=2p$.

The computation is straightforward as long as the saddle point
is unaffected. This is the case, for example, when there is a Wilson
line on a single gauge group, which gives \eqref{agreementf}-\eqref{agreements}.
Similarly,
when all the representation are anti-symmetric
we have the leading contribution
\begin{align}
 \langle W_{A_{k_1}, A_{k_2}, \ldots, A_{k_{q}}} \rangle &  = \exp \left[4 \pi
 \sqrt{\frac{n N}{2(8-N_f)}}N\sum_{a=1}^{q}
 \left(1-\left|1-\frac{k_a}{N}\right|^{3/2}\right)\right] \ .
\label{antisymmquiverCFT}
\end{align}
The result \eqref{antisymmquiverCFT} is simply a product of contributions from the Wilson loops located at
each gauge node. We will come back to the holographic interpretation of
this result later.

\begin{figure}[htbp]
\centering{\includegraphics[scale=0.6,trim=0 140 0 150]{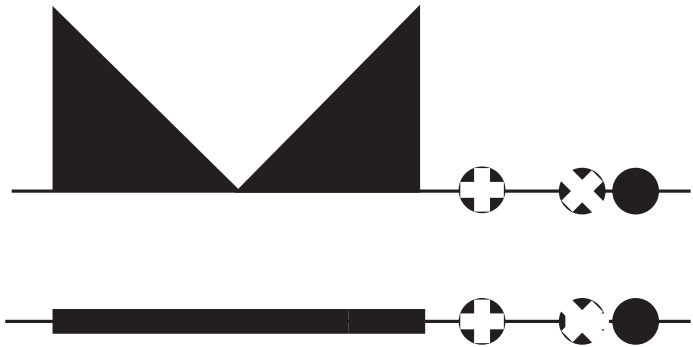}}
\caption{Wilson lines in symmetric representations for the quiver theories ($n>1$) can be described
by  excitations of interacting particles above the Fermi sea.
The size of the Fermi sea scales as $\sqrt{n}$.
There are several different types of particle species, corresponding to different nodes of
 the quiver.
In this figure the different particle species are represented by different types of dots.}
\label{fig.quiversea}
\end{figure}

The case with symmetric representations, however, is more subtle.
Let us consider $\langle W_{S_{k_1}, S_{k_2}, \ldots, S_{k_q}}\rangle $,
for example. By the same logic as in section \ref{CFT.symm},
we find that the dominant contribution is from the large winding Wilson loops.
This means that the matrix model reduces to an integral over the
eigenvalues $\sigma_1^{(a)}$, while all other eigenvalues can be replaced
by the smooth eigenvalue distribution.
The resulting effective matrix model is similar to \eqref{sigma1int},
however this time $\sigma_1^{(a)}$ with different values of $a$ interacts
among themselves, and a careful analysis is required.
In other words, for the $n>1$ case there are several different species of particle-like
excitations above the Fermi sea, and there are non-trivial
 interactions between them (Figure \ref{fig.quiversea}).

Since the general case is notationally involved,
let us first study the simplest non-trivial case of $n=2$
with gauge groups $USp(2N)\times USp(2N)$.
We parametrize the Cartan of the two gauge groups by $\rho_i, \sigma_i$
with $i=1, \ldots, N$.
In this case the integrand of the matrix model is $e^{-F[\rho, \sigma]}$
with
\begin{equation}
\begin{split}
F[\rho, \sigma]&=
\sum_{i\ne j}
\left[
  F_V(\rho_i+\rho_j)+F_V(\rho_i-\rho_j)
+F_V(\sigma_i+\sigma_j)+F_V(\sigma_i-\sigma_j)
\right] \\
&\quad + 2 \sum_{i}
\left[
F_V(2\rho_i)+F_V(2\sigma_i)
\right]
+2\sum_{i,j} \left[
F_H(\rho_i+\sigma_j)+F_H(\rho_i-\sigma_j)
\right] \\
&\quad +2\sum_i \left[
N_f^{(1)}
  F_H(\rho_i)
+
N_f^{(2)}
  F_H(\sigma_i)
\right] \ .
\end{split}
\end{equation}
The leading contribution to the symmetric Wilson loop
$\langle W_{S_k, S_l}\rangle$
comes from large winding modes,
contributing $e^{2\pi (k \rho_1+l \sigma_2)}$ to the matrix model.
This is justified by arguments similar to the $n=1$ case.
By replacing $\rho_i, \sigma_i, (i\ge 2)$ with smooth eigenvalue
distribution, we have an effective matrix model
\beq
\langle W_{S_{k}, S_{l}}\rangle
= \frac{
  \displaystyle \int d\sigma_1 \exp\left[-F_{\rm eff}(\rho_1, \sigma_1;k, l)
\right]
}
{
  \displaystyle \int d\sigma_1 \exp\left[-F_{\rm eff}(\rho_1, \sigma_1; 0,0)
\right]
} \ ,
\eeq
where we neglected the multiplicity factors which does not affect the
leading behavior, and defined
\begin{equation}
\begin{split}
& F_{\rm eff}(\rho_1, \sigma_1; k, l)=
-2\pi k \rho_1-2\pi l \sigma_1
+2 \left[F_H(\rho_1+\sigma_1)+F_H(\rho_1-\sigma_1)  \right] \\
& \quad + 2 \left[ F_V(2\rho_1)+F_V(2\sigma_1) \right]
 +2\left[
   N_f^{(1)} F_H(\rho_1)
+    N_f^{(2)} F_H(\sigma_1)
\right] \\
&\quad + 2N \int dy\, \rho(y)\left(
   F_V(\rho_1+N^{\half} y)+
  F_V(\rho_1-N^{\half} y)+
   F_V(\sigma_1+N^{\half} y)+
  F_V(\sigma_1-N^{\half} y)
\right)  \\
&\quad
+ 2N \int dy\, \rho(y)\left(
   F_H(\rho_1+N^{\half} y)+
  F_H(\rho_1-N^{\half} y)+
   F_H(\sigma_1+N^{\half} y)+
  F_H(\sigma_1-N^{\half} y)
\right)  \ . \\
\end{split}
\end{equation}
This contains terms of order $k N^{\half}$ and $N^{\frac{3}{2}}$ and lower;
terms of order $N^{\frac{5}{2}}$ cancel between $F_V$ and $F_H$.
Dropping terms of order $N^{\half}$, we obtain
\begin{equation}
\begin{split}
& F_{\rm eff}(x_1, x_2; k, l)=
 \pi x_0 N^{\frac{3}{2}}
\Bigg[
   -x_1\left(2\frac{k}{N}+\frac{9}{2}\right)
-x_2\left(2\frac{l}{N}+\frac{9}{2}\right) \\
 &\qquad
 +\frac{1}{3}x_0^2 (8- N_f^{(1)}) x_1^3
+\frac{1}{3}x_0^2 (8- N_f^{(2)}) x_2^3
-\frac{1}{3}x_0^2 (|x_1+x_2|^3+|x_1-x_2|^3)
\Bigg] \ ,
\end{split}
\label{Feff2}
\end{equation}
where we defined $x_1 = N^{-1/2}x_0^{-1} \sigma_1$, $x_2 = N^{-1/2}x_0^{-1} \rho_1$ and we assumed $x_1, x_2 > 0$.

Carrying the same analysis for arbitrary even $n=2p$ with symmetric
representation orders $(k_1,k_2,...,k_{p+1})$  (still considering the
case with vector structure) leads to the generalization of (\ref{Feff2})
\begin{equation}
\begin{split}
 F_{\rm eff}(x_a ; k_a) =
 \pi x_0 N^{\frac{3}{2}}
 & \Bigg[ \sum_{a=1}^{p+1}  \Big(
   -x_a \left(2\frac{k_a}{N}+\frac{9}{2} c_a \right)   +\frac{1}{3}x_0^2 (8- N_f^{(a)}) x_a^3  \Big)\\
 &\qquad\qquad -  \sum_{a=1}^{p} \frac{1}{3}x_0^2 \Big( |x_a-x_{a+1}|^3 + |x_a+x_{a+1}|^3  \Big)
\Bigg] \ ,
\end{split}
\label{Feffquiver}
\end{equation}
where $c_a=1$ ($c_a=2$) when the $a$-th gauge group is $USp(2N)$ ($SU(2N)$). \\
Let us here assume that flavors are distributed evenly, i.e. $N_f^{(a)}$
are the same for all $a$.
Let us moreover assume that $k_a/N+(9/4)c_a$ are the same for all $a$;
when $k_a\gg N$ this simplify means that $k_a$'s are the same
for all $a$.
Then by symmetry considerations it is easy to see that
there are saddle points at the locus
$x_a= x_1$ for all $a$.
This ansatz kills almost all the cubic terms and we are left with
\begin{equation}
\begin{split}
 F_{\rm eff}(x_1 ; k_a) =
 \pi x_0 N^{\frac{3}{2}}
 & \Bigg[
   -x_1 \left(2\frac{\sum_{a=1}^{p+1} k_a}{N}+\frac{9}{2} n \right)   + \frac{1}{3}x_0^2 \Big( 8- \sum_{a=1}^{p+1} N_f^{(a)} \Big) x_1^3
\Bigg] \ ,
\end{split}
\end{equation}
where we used the relation $\sum_{a=1}^{p+1} c_a = n$.
With $k_{\rm tot}:= \sum_{a=1}^{p+1} k_a$, $N_f =  \sum_{a=1}^{p+1} N_f^{(a)}$ and the rescaling $x_1 = n^{1/2} \tilde x_1$, we obtain
\begin{equation}
\begin{split}
 F_{\rm eff}(\tilde x_1 ; k_{\rm tot}) =
 \pi x_0 n^{\frac{3}{2}} N^{\frac{3}{2}}
 & \Bigg[
   -\tilde x_1 \left(2\frac{k_{\rm tot}}{n N}+\frac{9}{2}  \right)   + \frac{3}{2} \tilde x_1^3
\Bigg] \ .
\end{split}
\end{equation}
This brings us back to the non-orbifold case (\ref{Feff}) with a
$n^{3/2}$ prefactor and the replacement $k \rightarrow k_{\rm tot}/n$.

We thus obtain in the end
\beq
\langle W_{S_1, \ldots, S_q}\rangle  =
\exp\left[
  \frac{9\pi }{\sqrt{2(8-N_f)}}n^{\frac{3}{2}} N^{\frac{3}{2}} \left[
\left(1+\frac{4k_{\rm tot}}{9n N}\right)^{\frac{3}{2}}-1
\right]
\right] \ .
\label{quiversymmresultCFT}
\eeq
As we will see in section \ref{symD4},
this result matches with the holographic computation.

The same analysis can be done for odd $n$ and even $n$ without vector structure and it leads, under the same assumptions, to the same result (\ref{quiversymmresultCFT}).

The more general cases, when the $k_a/N+(9/4)c_a$ are different and the numbers of flavor $N^{(a)}_f$ in each node are different, are more involved. The $x_a$ at the saddle point are no longer equal. In this case, a match with gravity computations would require a more complete description of the type IIA background, for example, by including discrete holonomies of the B-field on the 2-cycles of the geometry and restoring the dependence on the $N^{(a)}_f$ parameters.

\section{Holographic Computations} \label{sec.gravity}

In this section we reproduce the same results from the holographic
computation in the dual massive IIA supergravity background
\cite{Brandhuber:1999np,Bergman:2012kr} (see also \cite{Passias:2012vp}
for uniqueness).
First we review the solution.  The metric is given by
\begin{align}
\label{metric}
ds^2 = \frac{L^2}{(\sin \alpha)^\frac{1}{3}} \left[ ds^2_{AdS_6} +
 \frac{4}{9} \left(d\alpha^2 + \cos^2 \alpha \, ds^2_{S^3/\mathbb{Z}_n}
 \right) \right],
\end{align}
where $\alpha$ ranges as $\alpha \in (0,\pi/2]$.  The orbifold is realized by writing the $S^3$ metric as
\begin{align}
ds^2_{S^3/\mathbb{Z}_n} = \frac{1}{4} \left[ d \theta_1^2 + \sin^2
 \theta_1 d\theta_2^2 + (d\theta_3 - \cos \theta_1 d\theta_2)^2 \right],
\end{align}
and taking the angles to range as $\theta_1 \in [0,\pi)$, $\theta_2 \in [0,2
\pi)$ and $\theta_3 \in [0,4 \pi/n)$.
The AdS radius $L$ is related to the integer parameters $n, N, N_f$ by
\begin{align}
\label{radius}
\frac{L^4}{l_s^4} = \frac{18 \pi^2 n N}{8-N_f} \ .
\end{align}
The dilaton $\phi$, and Roman's mass $F_0$, are given by
\begin{align}
&e^{-2 \phi} = \frac{3(8-N_f)^\frac{3}{2} (n
 N)^{\frac{1}{2}}}{2^\frac{3}{2} \pi} \sin^\frac{5}{3} \alpha \ ,&
&F_{(0)} = \frac{8 - N_f}{2 \pi l_s} \ .&
\end{align}
Note that the dilaton diverges at $\alpha=0$.
Near this region the curvature also diverges, and
the supergravity approximation breaks down.

There is also a 6-form flux corresponding to the presence of D4-branes
\begin{align}
F_{(6)} = - 45 \pi n N L^2 l_s^3 \, \omega_{AdS_6} \ ,
\end{align}
where $\omega_{AdS_6}$ is the unit volume form on $AdS_6$.  The number of D4-branes this flux corresponds to can be computed as follows.  First we compute $F_{(4)} = * F_{(6)}$
\begin{align}
F_{(4)} =& 45 \pi n N l_s^3 \left(\frac{2}{3}\right)^4 (\sin^\frac{1}{3}
 \alpha \cos^3 \alpha) d \alpha \wedge \omega_{S^3/\mathbb{Z}_n} \ ,
\end{align}
where $\omega_{S^3/\mathbb{Z}_n} = (\sin(\theta_1)/8) d\theta_1 \wedge d\theta_2 \wedge d\theta_3$. Integrating to get the charge, we have
\begin{align}
Q_{\rm D4} = \frac{1}{2 \kappa^2} \int F_{(4)} = T_4 N \ ,
\end{align}
where we have used $2 \kappa^2 = (2 \pi)^7 l_s^8$ and the D4-brane tension is $T_4 = 1 / [(2\pi)^4 l_s^5]$.
To help with computations, we introduce the notations
\begin{align}
e^{-2 \phi_0} :=& \frac{3(8-N_f)^\frac{3}{2} (n
 N)^{\frac{1}{2}}}{2^\frac{3}{2} \pi } \ ,&
Q_4 :=& 45 \pi n N l_s^3 \left(\frac{2}{3}\right)^4 \ ,&
Q_6 :=& - 45 \pi n N L^2 l_s^3\ .&
\end{align}

This solution preserves 16 supersymmetries.  As discussed in Appendix \ref{app:susy}, the ten-dimensional supersymmetry parameter can be decomposed into a basis of Killing spinors as follows
\begin{align}
\epsilon = \sum_{\eta = \pm} \tilde \chi^{(2)}_{\eta} \otimes \tilde \chi^{(3)}_{\eta} \otimes \tilde \zeta_{\eta,-\eta} \otimes \zeta_{+,+} \otimes \chi_{+}^{(3)} \ ,
\end{align}
where $\tilde \chi^{(2)}_{\eta}$ is a Killing spinor on $AdS_2$, $\tilde \chi^{(3)}_{\eta}$ is a Killing spinor on $S^3$ and $\chi_{+}^{(3)}$ is a Killing spinor on $S^3$.  The remaining components $\zeta_{\eta,-\eta}$ and $\zeta_{+,+}$ satisfy the projection conditions
\begin{align}
\label{susyproj}
\zeta_{+,+} =& \left[\cos(\alpha) \sigma^2 + \sin(\alpha) \sigma^1 \right] \zeta_{+,+} \ , \cr
\tilde \zeta_{\eta,-\eta} =& \eta \left[ i \sigma^2 \sinh(x) + \sigma^1 \cosh(x) \right] \tilde \zeta_{\eta,-\eta} \ .
\end{align}
Additionally, each Killing spinor and $\zeta_{\eta,-\eta}$ and $\zeta_{+,+}$ satisfy reality conditions.  Note that the combination $\chi_{+}^{(6)} = \sum_{\eta = \pm} \tilde \chi^{(2)}_{\eta} \otimes \tilde \chi^{(3)}_{\eta} \otimes \tilde \zeta_{\eta,-\eta}$ yields a Killing spinor on $AdS_6$.  Counting degrees of freedom, we have $8 \times 2 = 16$ parameters.  The $2$ comes from $\chi_{+}^{(3)}$ while the $8$ comes from $\chi_{+}^{(6)}$.

The gravitational dual of the Wilson loop \eqref{loopdef} in the fundamental representation
is the fundamental string \cite{Rey:1998ik,Maldacena:1998im}. However, when we consider
general anti-symmetric or symmetric representations, the fundamental string is replaced by
D-branes (for the similar case of D3-branes see \cite{Drukker:2005kx,Yamaguchi:2006tq,Gomis:2006sb}).
There are two possibilities:
\begin{enumerate}[label=({\arabic*})]
\item D4-branes wrapping AdS$_2$ and an internal $S^3$,
\item D4-branes wrapping AdS$_2$ and the space-time $S^3$.
\end{enumerate}
Roughly, they respectively correspond to anti-symmetric and symmetric representations.  To be more precise, they should be dual to irreducible representations.  For $USp(2N)$ groups the anti-symmetric representations are reducible, as discussed in Appendix \ref{sec.grouptheory}, and the D4-branes are dual to the largest irreducible component of the anti-symmetric representations.
The corresponding flat space brane configuration is given in Table \ref{table.branes}.

This identification can be motivated as follows.  First, these two D4-branes are the only branes which preserve the same $\half$-BPS
supersymmetry as preserved by the fundamental string (Table
\ref{table.branes}).\footnote{In general the branes impose projections on the supersymmetry parameters.  For each brane we introduce, there is a quantity $\delta_i$ which imposes the constraint $\delta_i \varepsilon = \varepsilon$.  Generically the $\delta_i$ are traceless matrices with eigenvalues $\pm 1$.  In order for the projection operators to be compatible, the $\delta_i$ must commute.  The explicit $\delta_i$ for the above branes are given by
$\delta_{O8^-/D8} = \Gamma^9 \Gamma^{\sharp}$, $\delta_{\rm D4}
=\delta_{{\rm D4}_{\rm  symm}}=
\Gamma^{56789} \Gamma^{\sharp}$, $\delta_{{\rm D4}_{\rm antisymm}} = \Gamma^{12349}
\Gamma^{\sharp}$.}
Secondly, if we consider fundamental strings stretched between the background stack of branes and the stack of D4-branes in case (1), we find the number of Dirichlet-Neumann directions is 8.  This means the zero energy ground state of such strings is in the R sector in the NSR
formalism.  This behaves as a fermion and hence anti-symmetrizes the Chan-Paton indices, so that the fundamental strings are naturally anti-symmetrized.  Similarly, for case (2) the ground state is in the NS sector, and correspondingly we end up with symmetric representations.

\begin{table}[htbp]
\begin{centering}
\begin{tabular}{|c||c||c|c|c|c||c|c|c|c||c|}
  \hline
     & 0 & 1 & 2 & 3 & 4 & 5 & 6 & 7 & 8 & 9 \\ \hline
  O8$^-$/D8 & X & X & X & X & X & X & X & X & X &  \\
  D4 & X & X & X & X & X &   &   &   &   &   \\ \hline
  F1 & X &  &  &  &  &   &   &   &   & X  \\
  D4$_{\rm antisymm}$ & X &   &   &   &   & X & X & X & X &   \\
  D4$_{\rm symm}$ & X & X & X & X & X &   &   &   &   &   \\
  \hline
\end{tabular}
\caption{Supersymmetric brane configurations. The two types of
 D4-branes, D4$_{\rm symm}$ and D4$_{\rm antisymm}$, preserve the same
 $\half$-BPS supersymmetry as preserved by the fundamental string.}
\label{table.branes}
\end{centering}
\end{table}

Let us comment more on the orbifold case $n>1$.
The orbifold $\bZ_n$ does not have a fixed point on $S^3/\bZ_n$, however
it does have a fixed point on the 4d space spanned by $\alpha, \theta_1,
\theta_2, \theta_3$.  Locally near $\alpha=\frac{\pi}{2}$,
we have an orbifold singularity of the form $\bC^2/\bZ_n$.  Correspondingly, the geometry contains additional
2-cycles associated to the twisted sectors.
The (probe) D-branes wrapping different 2-cycles will correspond to Wilson loops in representations of different gauge nodes in the quiver theory.

\begin{figure}[htbp]
\centering{\includegraphics[scale=0.4,trim=0 50 0 50]{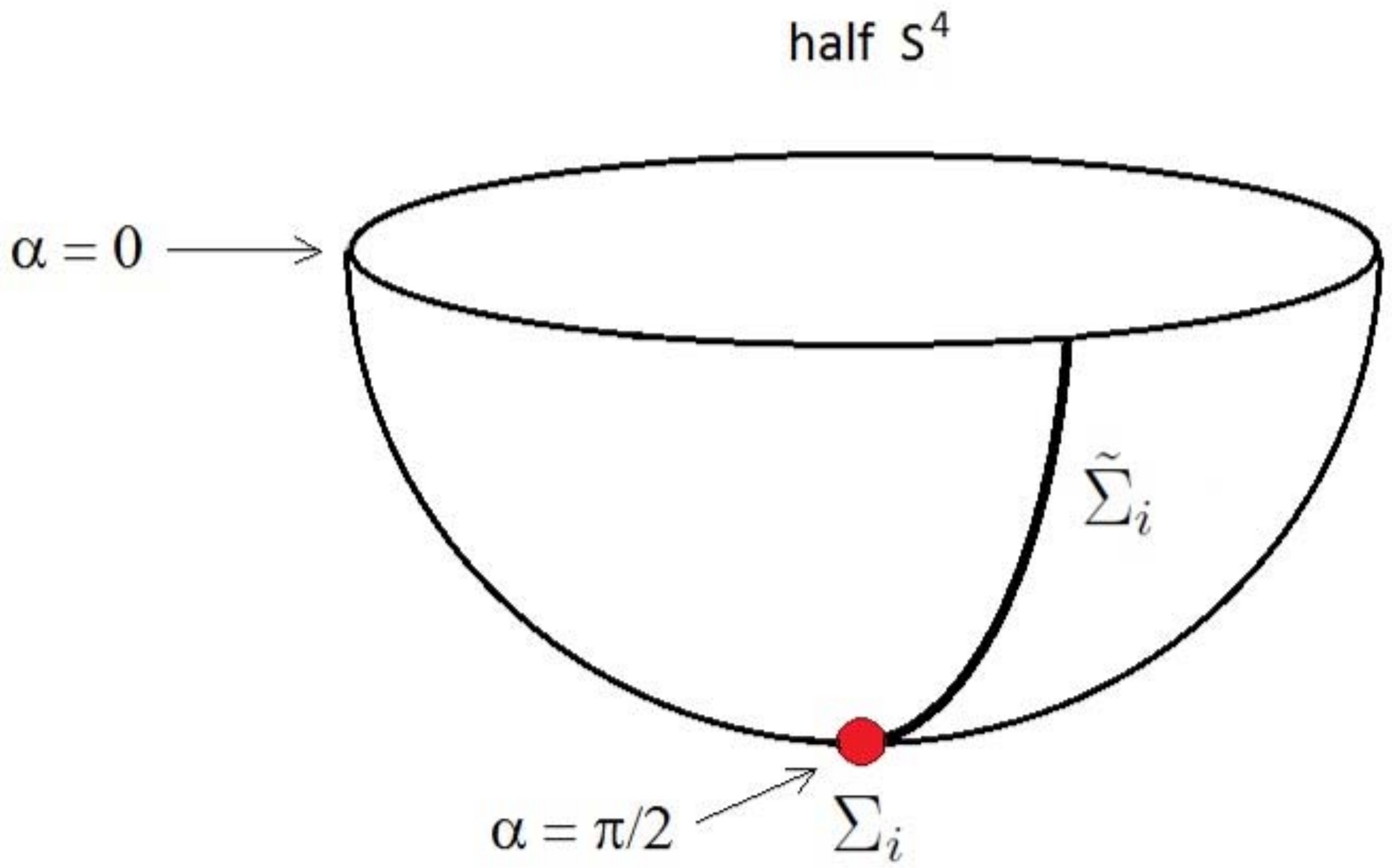}}
\caption{Vanishing 2-cycles $\Sigma_i$ at the pole of the semi-$S^4$ and dual 2-cycles $\tilde \Sigma_i$ ($1\le i \le n-1$) spanned by coordinates $\alpha$ and $\theta_1$ ($S^1$ transverse to the picture). }
\label{2cycles}
\end{figure}

As detailed in \cite{Bergman:2012kr} the various 2-cycles can be seen in
the resolved geometry where the $\bZ_n$ orbifold is blown up to a
$n$-centered ALE space.  The orbifold corresponds to the
limit when all centers merge to the same point.
More precisely in our IIA geometry, there are $n-1$ vanishing 2-cycles
$\Sigma_i$ wrapping the coordinates $\theta_1, \theta_2$ at the orbifold
singularity $\alpha = \pi/2$ and $n-1$ dual 2-cycles $\tilde{\Sigma}_i$
wrapping the coordinates $\alpha, \theta_3$ (Figure
\ref{2cycles}). However not all of these cycles are independent; the orientifold projection maps the $i$-th twisted sector with the ($n-i$)-th twisted sector, this implies that the $\Sigma_i$ ($\tilde{\Sigma}_i$) should be identify with $\Sigma_{n-i} (\tilde{\Sigma}_{n-i})$.  
The branes wrapping these cycles have the following interpretation.

\begin{figure}[htbp]
\centering{\includegraphics[scale=0.4,trim=0 50 0 50]{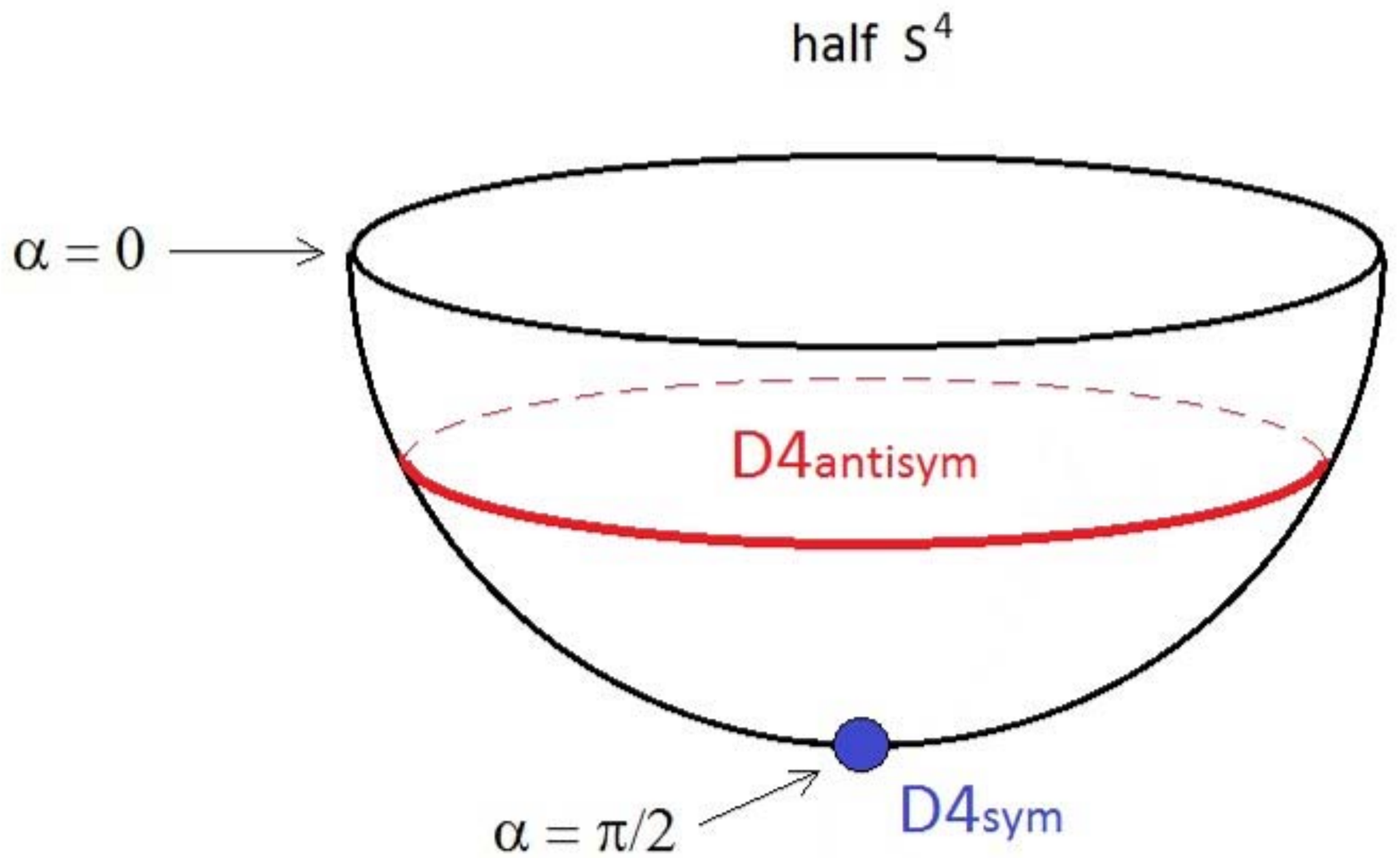}}
\caption{D4-branes embeddings in the internal space for symmetric and
 antisymmetric cases. The D4$_{\rm antisymm}$-brane wraps the fibered $S^2$ ($\theta_1,\theta_2$) in red and the $S^1$ ($\theta_3$) transverse to the picture.}
\label{D4embed}
\end{figure}

Wilson loops in (the largest irreducible component of) the anti-symmetric representations of one of the quiver nodes are dual to D4-branes wrapping an $AdS_2\times \tilde S^3_i$,  with
the 3-sphere $\tilde S^3_i= \tilde S^1_i \times S^2$, where $S^2$ is
the 2-sphere parametrized by $\theta_1, \theta_2$ and $\tilde S^1_i$ is
the circle in $\tilde \Sigma_i$ parametrized by $\theta_3$ (Figure \ref{D4embed}).  
To support this picture, we match the number of D4-brane embeddings to the number of Wilson loops as follows.
When $n$ is odd, there are $[n/2]$ such additional $\tilde S^3_i$ cycles.  The D4-branes wrapping these cycles correspond to anti-symmetric representations of the $SU(2N)$ gauge groups, while the D4-brane wrapping the original cycle corresponds to a representation of $USp(2N)$.  When $n$ is even, there is a cycle which is mapped into itself under the orientifold projection.  There are then two cases to consider.  In either case, there are $[n/2]-1$ $\tilde S^3_i$ cycles which the D4-branes can wrap yielding anti-symmetric representations of $SU(2N)$ gauge groups.  In the case with vector structure, the D4-branes wrapping the remaining $\tilde S^3_{[n/2]}$ cycle and the original cycle yield representations of the two remaining $USp(2N)$ gauge groups.  Finally in the case without vector structure, the D4-branes wrapping the remaining $\tilde S^3_{[n/2]}$ cycle and the original cycle must combine to yield representations of $SU(2N)$.

For Wilson loops in symmetric representations of one of the gauge factors,
the holographic dual is a D4-brane wrapping the spacetime $AdS_2\times
S^3$ and sitting at the point $\alpha = \pi/2$ in internal space (Figure
\ref{D4embed}).  To obtain such a configuration we can either have a true D4-brane sitting at $\alpha = \pi/2$ or D6-branes with the same space-time
embedding wrapped on the vanishing two-cycle $\Sigma_i$ at $\alpha =
\pi/2$.\footnote{This is analogous to the fractional D4-branes that
are D6-branes wrapped on $\Sigma_i$ and which increase the rank of the
corresponding gauge factor in the quiver theory. However to determine
the gravity duals of quivers with factors of different ranks, one
should take into account the backreaction of the fractional
D4-branes.}

When we have Wilson loops in non-trivial representations for several gauge groups, we might expect the dual to be described by several D-branes wrapping distinct cycles.  For anti-symmetric representations, we have several D4-branes parallel to each other at different values of
$\alpha$ (see Figure \ref{D4s}), and this picture is naturally realized in the CFT result \eqref{antisymmquiver}.
For symmetric representations, we have seen that the factorization does not hold.
This unexpected result could be related to the fact that the D4-branes corresponding to symmetric representations
are really fractional D6-branes wrapping different 2-cycles whose size is vanishing in the orbifold limit.
In this case, there are additional contributions one might have to take into account, for example, from discrete holonomies of the B-field on these 2-cycles.  Additionally, one needs to be able to account for the $N^{(i)}_f$ dependence appearing in the quiver gauge theories.

In the rest of this section  we compute the worlsheet action of the single fundamental string and the worldvolume actions of D4-branes in the anti-symmetric and symmetric embeddings.  We find perfect agreement with the matrix model computations of the previous section. For completeness we also evaluate the IIA action on the supergravity solution and match it with the free energy computation on the 5-sphere.

\subsection{Fundamental Representation}

The Wilson loops preserve the bosonic symmetry $SO(1,2) \times SO(4)
\times SU(2)_R \times SU(2)_M$, when $n=1$ (section \ref{sec.CFT}).  We will therefore choose coordinates on $AdS_6$, which make this symmetry manifest
\begin{align}
\label{AdSslice}
ds^2_{AdS_6} =& \cosh^2(x) ds^2_{AdS_2} + \sinh^2(x) ds^2_{S^3} + dx^2,
\end{align}
where for the symmetric spaces, we choose the coordinates (see Appendix \ref{app:embed})
\begin{align}
ds^2_{AdS_2} =& \frac{1}{\sinh^2 \rho} (d\rho^2 - d\psi^2) \ , \cr
ds^2_{S^3} =& d \phi_1^2 + \sin^2(\phi_1) d \phi_2^2 + \sin^2(\phi_1)
 \sin^2(\phi_2) d \phi_3^2 \ .
\end{align}
When $n>1$, the Wilson loops are not charged under the broken $SU(2)_M$ symmetry and the brane embeddings for the $n=1$ case map, in straightforward way, to brane embeddings in the $n>1$ case.  In the following we will allow for general values of $n$.

We first consider a fundamental string with world-volume coordinates $\xi^i$ with $i = 0,1$.  We take the fundamental string to wrap the $AdS_2 $ slice.  In order to preserve the $SO(4) \times SU(2)_R \times SU(2)_M$ symmetry, the string must sit at locations where the two $S^3$'s vanish, namely at $x = 0$ and $\alpha = \pi/2$.  One can check that this choice is in fact an extremum of the Nambu-Goto action.  Denoting the induced metric as $G_{ij}$, the on-shell action is given by
\begin{align}
S_{\rm F1} = - \frac{1}{2 \pi l_s^2} \int d\xi^i \sqrt{-\det(G_{ij})}
= - \frac{3 \sqrt{2 n N}}{2 \sqrt{8 - N_f}} \int d\rho d\psi \,
 \frac{1}{\sinh^2(\rho)} \ .
\end{align}
This answer is divergent even after taking $\psi$ to be compact.  To get a finite answer, we compute the Legendre transformed action\footnote{Alternatively, one can use holographic renormalization, including counterterms, to arrive at the same result.}.  The reason for the Legendre transform is that the dual of a supersymmetric Wilson loop is a fundamental string which satisfies Dirichlet boundary conditions parallel to the boundary and Neumann boundary conditions perpendicular to the boundary \cite{Drukker:1999zq}.  For our simple string, we take $\xi^0 = \psi$ and keep $\xi^1$ arbitrary.  The profile of the string is then given by $z = \xi^1$ and the Legendre transformed action is given by
\begin{align}
A_{\rm F1} = S_{\rm F1} -\int d\psi  \left( z \frac{\p S_{\rm F1}}{\p (\p_{\xi^1} z)}  \right) \bigg|_{z = 0}
= \frac{3 \sqrt{2 n N}}{2 \sqrt{8 - N_f}} (2 \pi R_{\psi}) \ ,
\end{align}
where we have taken the $\psi$ direction to be compact with periodicity $\psi = \psi + 2 \pi R_{\psi}$.
Going to the Euclidean, we set $R_{\psi} = 1$, as discussed in
Appendix \ref{app:embed}.  We then arrive at the advertised result \eqref{agreementf} for $n=1$ and \eqref{fundorb} for $n>1$.

Next we check the supersymmetry of the embedding.  The projection corresponding to the fundamental string is given by
\begin{align}
\label{projfund}
\epsilon = \pm \Gamma^{01} \Gamma^\sharp \epsilon = \pm \Gamma^{23456789} \epsilon \ ,
\end{align}
where one chooses a definite sign.  Using our conventions given in Appendix \ref{app:susy} we have
\begin{align}
\Gamma^{23456789} = \id_4 \otimes \sigma^1 \otimes \sigma^1 \otimes \id_2 \ .
\end{align}
This is compatible with the projections \eqref{susyproj} on $\zeta_{+,+}$ and $\tilde \zeta_{\eta,-\eta}$ provided $x = 0$, $\alpha = \pi/2$ and $\eta = \pm 1$, where the sign choice is correlated with the choice in \eqref{projfund}. The restriction of $\eta$ to a definite sign reduces the number of supersymmetries by half.

\subsection{Anti-symmetric Representations}\label{antiD4}

We consider a D4-brane with world-volume coordinates $\xi^i$ with $i = 0,...,4$.  We take the D4-brane to wrap the internal $S^3$ and the $AdS_2 $ slice.  In this case we can make the identification
\begin{align}
&\xi^0 = \psi,& &\xi^1 = \rho,& &\xi^2 = \theta_1,& &\xi^3 = \theta_2,&
 &\xi^4 = \theta_3 \ .&
\end{align}
We take a worldvolume flux proportional to the $AdS_2$ volume
\begin{align}
{\cal F} = q L^2 \frac{\cosh^2 (x)}{\sin^{\frac{1}{3}}(\alpha) \sinh^2
 (\rho)} d \rho \wedge d \psi \ ,
\end{align}
where $q$ is an arbitrary coefficient, which can depend on both $\alpha$ and $x$.
It will be necessary to have an explicit expression for $C_{(3)}$
\begin{align}
C_{(3)} = Q_4  \frac{3}{40} \sin^\frac{1}{3}(\alpha) [ 7 \sin(\alpha) + \sin(3 \alpha) ] \omega_{S^3/\mathbb{Z}_n} - Q_4 \frac{18}{40} \omega_{S^3/\mathbb{Z}_n} \ .
\end{align}
Note that the choice of $C_{(3)}$ is not unique and in particular one
can make large gauge transformations which are proportional to the unit
volume form on the $S^3/\mathbb{Z}_n$.  However, we note that the
$S^3/\mathbb{Z}_n$ is a
vanishing cycle at $\alpha = 0$.  In order for $C_{(3)}$ to be regular, we should then require $C_{(3)}$ to vanish at $\alpha = 0$, which then fixes the gauge freedom as above.

In order to preserve the symmetry of the space-time $S^3$, the D4-brane must sit at $x=0$.  It is then consistent to take the remaining embedding coordinates, namely $\alpha$, to be constant.  One can check that this satisfies the general equations derived in \cite{Skenderis:2002vf}.
Introducing the induced metric $G_{ij}$ and the pullback of $C_{(3)}$ as $\hat C_{(3)}$, the D4-brane action is given by
\begin{align}
S_{\rm D4} =& - T_4 \int d^5\xi \, e^{-\phi} \sqrt{ -\det(G_{ij} + {\cal F}_{ij})} + T_4 \int {\cal F} \wedge \hat C_{(3)} \cr
=& - T_4 \text{vol}(S^3/\mathbb{Z}_n) \int d\rho d\psi \,  e^{-\phi_0} \frac{L^5}{\sinh^2(\rho)} \left(1 - \frac{\sinh^4(\rho)\sin^\frac{2}{3}(\alpha)}{L^4} ({\cal F}_{\rho \psi})^2 \right)^\frac{1}{2} \left(\frac{2 \cos(\alpha)}{3} \right)^3 \cr
&- T_4 \text{vol}(S^3/\mathbb{Z}_n) \int d\rho d\psi \, {\cal F}_{\rho
 \psi} Q_4 \left[ \frac{3}{40} \sin^\frac{1}{3}(\alpha) \left( 7
 \sin(\alpha) + \sin(3 \alpha) \right) - \frac{18}{40} \right] \ .
\end{align}
We take $d \psi \wedge d \rho \wedge \omega_{S^3/\mathbb{Z}_n}$ to be positive, which accounts for the sign in the second term.
Minimizing the above action for $\alpha$ with fixed ${\cal F}$ and then plugging in the expression for ${\cal F}$ yields an equation which determines $\alpha$
\begin{align}
- 81 q \sqrt{1-q^2} Q_4 \cos(\alpha) \sin(\alpha) + 8 e^{-\phi_0} L^3
 \left[q^2(1 - 10 \sin^2(\alpha)) + 9 \sin^2(\alpha) \right] = 0 \ .
\end{align}
Of course the above procedure is not necessarily consistent and we have checked that this equation can also be obtained using the general equations derived in \cite{Skenderis:2002vf}.  Plugging in the explicit expressions for $Q_4$, $\phi_0$ and $L$ leads to
\begin{align}
(9-10 q^2) \sin^2(\alpha) + q(q-5 \sqrt{1-q^2}) \sin(2 \alpha) = 0 \ .
\end{align}
This can be solved to give $q$ in terms of $\alpha$.  There are two solutions
\begin{align}
\label{qsolanti}
&q = \pm \sin(\alpha) \ ,&
&q = \pm 9 \sqrt{\frac{1 - \cos(2\alpha)}{82-80 \cos(2\alpha)}} \ .&
\end{align}
The first solution is compatible with supersymmetry while the second is not.  We therefore consider only the first solution.
The quantization condition of the fundamental string charge
is given in \eqref{F1chargeant}
\begin{align}
N_{\rm F1} = N - N \frac{\sin^\frac{1}{3}(\alpha)}{6} \left(\sin(3 \alpha) + 7 \sin(\alpha) - \frac{4q}{\sqrt{1-q^2}} \cos^3(\alpha) \right),
\end{align}
with $N_{\rm F1}$ the number of fundamental strings dissolved into the D4-brane.
After plugging in the expression for $q$, we obtain an expression giving
$N_{\rm F1}$ in terms of $\alpha$
\begin{align}
N_{\rm F1} =& N - N \sin^\frac{4}{3}(\alpha) \ .
\end{align}
Note this solution satisfies $N_{\rm F1} = 0$ when $\alpha = \pi/2$ and $N_{\rm F1} = N$ when $\alpha = 0$.  This is consistent with the matching of these D4-brane embeddings to anti-symmetric representations.

Computing the on-shell action, we find
\begin{align}
\label{gravanti}
S_{\rm D4} =& - T_4 \int d^5\xi \, e^{-\phi} \sqrt{ -\det(G_{ij} + {\cal F}_{ij})} + T_4 \int {\cal F} \wedge \hat C_{(3)} - \frac{N_{\rm F1}}{2 \pi l_s^2} \int {\cal F}
\cr
=& \frac{2 N}{3} \left[1 - \left(1 - \frac{N_{\rm F1}}{N}
 \right)^\frac{3}{2} \right] S_{\rm F1} \ .
\end{align}
The last term in the first line is a boundary term resulting from the coupling of the worldvolume gauge field to the boundary of the open string.
As a consistency check we remark that in the limit of small $N_{\rm F1}$ the position of the D4-brane goes to $\alpha = \pi/2$ where the internal $S^3$ vanishes and we recover the fundamental string wrapped on $AdS_2$, sitting at $(x,\alpha) = (0,\pi/2)$ as expected.

Surprisingly the result agrees with the gauge computation when $N_{\rm
F1} \rightarrow N$ ($k \rightarrow N$). In this limit the position of
the D4-brane is pushed to $\alpha = 0$ where the orientifold sits and we
might have expected that the supergravity background gets corrected in
this region. The reason why the holographic computation remains valid in
this region is unclear and deserves more attention.

\begin{figure}[htbp]
\centering{\includegraphics[scale=0.4,trim=0 50 0 0]{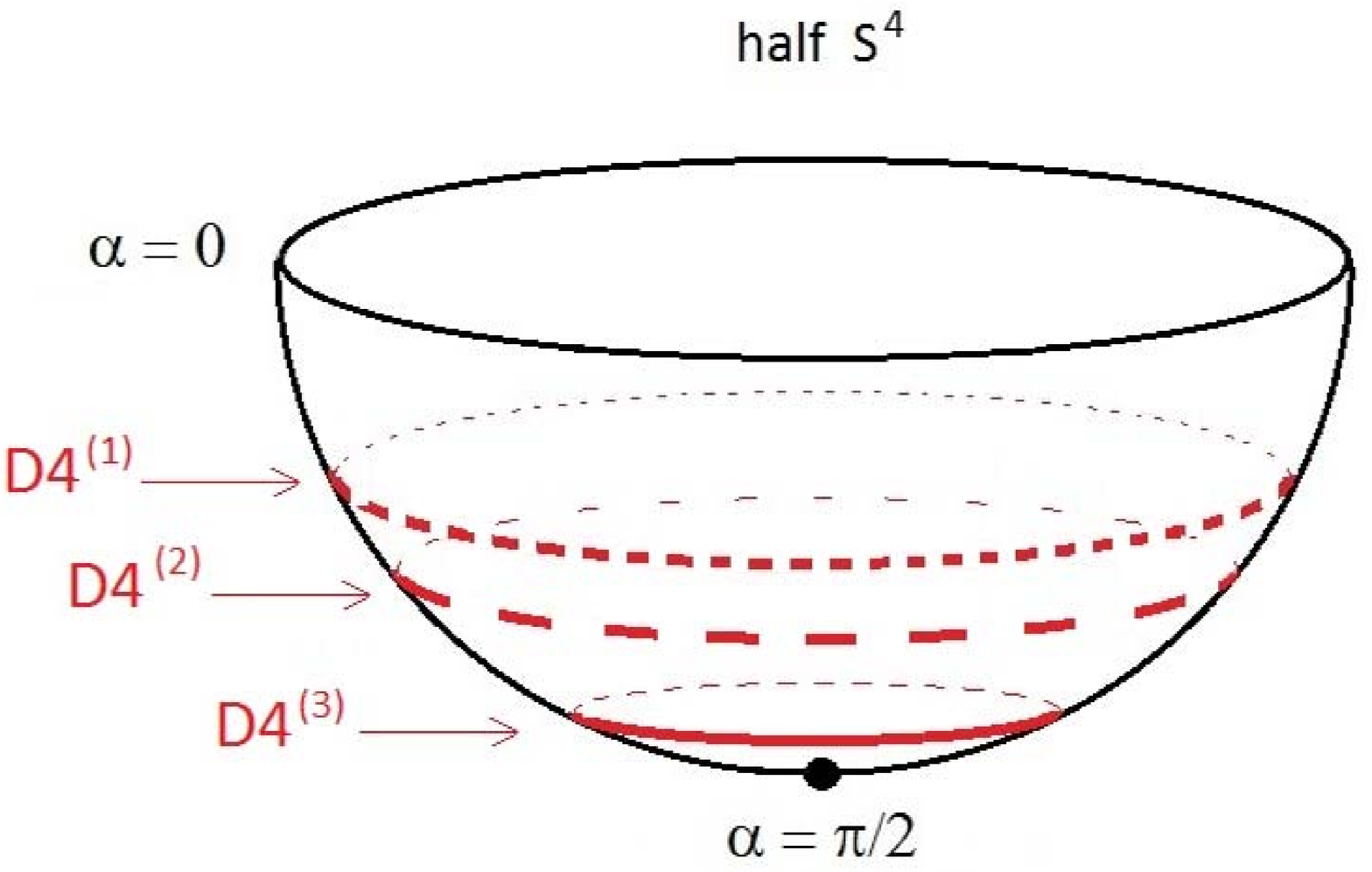}}
\caption{General antisymmetric representations for Wilson loops correspond to having D4-branes of different type (wrapping different cycles) at different positions in internal space.  Note that each line wraps a different $\tilde S^1_i$ cycle.}
\label{D4s}
\end{figure}

For $n=1$, \eqref{gravanti} gives the advertised result for anti-symmetric representations \eqref{agreementa}.
The result for more general representations, \eqref{generala}, can be interpreted as
the sum over contributions from multiple D4-branes, with one D4-brane for each $l_a$ in the representation sitting at
the position $\alpha_a$, as determined by the value of $l_a$.
Similarly, the result (\ref{antisymmquiver}) for anti-symmetric representations for Wilson loops in quiver theories ($n>1$) is obtained simply by adding contributions of multiple D4-brane actions, with each D4-brane sitting at a position $\alpha_a$ in the internal space determined by the order $k_a$ of the representation in the node $a$ (Figure \ref{D4s}).
However the important difference between the two is that in the latter case
the D4-branes are distinct in the sense that they wrap different $\tilde S^1_i$-cycles, as discussed at the beginning of this section.
If we consider general representations for quiver theories, we have in
general several D4-branes, sitting at different positions and on different cycles.

We now check the supersymmetry of this embedding.  The projection matrix \eqref{projmat} reduces to
\begin{align}
\Gamma =& \frac{1}{\sqrt{1-q^2}} \Gamma^{23456} - \frac{q}{\sqrt{1-q^2}} \Gamma^{789} \cr
=& -\frac{1}{\sqrt{1-q^2}} \left( \id_2 \otimes \id_2 \otimes \sigma^1 \otimes \sigma^2 \otimes \id_2 \right)
- \frac{q}{\sqrt{1-q^2}} \left( \id_2 \otimes \id_2 \otimes \id_2
 \otimes i \sigma^3 \otimes \id_2 \right) \, .
\end{align}
At $x=0$, the project condition \eqref{susyproj} on $\tilde \zeta_{\eta,-\eta}$ reduces to $\sigma^1 \tilde \zeta_{\eta,-\eta} = \eta \tilde \zeta_{\eta,-\eta}$.  The constraint $\epsilon = \Gamma \epsilon$ then reduces to
\begin{align}
\zeta_{+,+} = \eta \left( q \sigma^1 - \sqrt{1-q^2} \sigma^2 \right) \zeta_{+,+}
 \ .
\end{align}
This is compatible with \eqref{susyproj} for $\eta = -1$, provided we take $q = - \sin(\alpha)$.  One can easily check that taking the second solution in \eqref{qsolanti} yields a projection on $\zeta_{+,+}$ which is incompatible with \eqref{susyproj} and thus breaks all supersymmetries.

\subsection{Symmetric Representations}
\label{symD4}

We consider a D4-brane with world-volume coordinates $\xi^i$ with $i = 0,...,4$.  We take the D4-brane to wrap the space-time $S^3$ and the $AdS_2$ slice.  In this case we can make the identification
\begin{align}
&\xi^0 = \psi,& &\xi^1 = \rho,& &\xi^2 = \phi_1,& &\xi^3 = \phi_2,&
 &\xi^4 = \phi_3 \ .&
\end{align}
We again take a worldvolume flux proportional to the $AdS_2$ volume
\begin{align}
{\cal F} = q L^2 \frac{\cosh^2 (x)}{\sin^{\frac{1}{3}}(\alpha) \sinh^2
 (\rho)} d \rho \wedge d \psi  \ .
\end{align}
We will need the 5-from gauge potential.  In the coordinates \eqref{AdSslice}, $C_{(5)}$ is given by
\begin{align}
C_{(5)} = \frac{Q_6}{\sinh^2(\rho)}\! \left[ \frac{\cosh^3(x)}{30} \left(3 \cosh(2x) - 7\right) + \frac{4}{30} \right] \sin^2(\phi_1) \sin(\phi_2)d\rho \wedge d\psi \wedge d \phi_1 \wedge d \phi_2 \wedge d\phi_3 \ .
\end{align}
As before, we have the freedom to make large gauge transformations, which are proportional to the unit volume forms on the $AdS_2$ and $S^3$.  Since the $S^3$ vanishes at $x=0$, we require $C_{(5)}$ to vanish there as well, which fixes this gauge choice.

In order for the D4-brane to preserve the remaining $SO(4)$ symmetry of the internal $S^3$, it must sit at $\alpha = \pi/2$.  With this requirement, we can then again take the remaining embedding coordinates to be constant.  Again, one can check that this satisfies the general equations derived in \cite{Skenderis:2002vf}.  Introducing the induced metric $G_{ij}$ and the pullback of $C_{(5)}$ as $\hat C_{(5)}$, the D4-brane action is given by
\begin{align}
S_{\rm D4} =& - T_4 \int d^5\xi \,  e^{-\phi} \sqrt{ -\det(G_{ij} + {\cal F}_{ij})} + T_4 \int \hat C_{(5)} \cr
=& - T_4 \text{vol}(S^3) \int d\rho d\psi \, e^{-\phi_0} \frac{L^5}{\sinh^2(\rho)} \cosh^2(x) \sinh^3(x) \left(1 - \frac{\sinh^4(\rho)}{L^4 \cosh^4(x)} ({\cal F}_{\rho \psi})^2 \right)^\frac{1}{2} \cr
& - T_4 \text{vol}(S^3) \int d\rho d\psi \, \frac{Q_6}{\sinh^2(\rho)}
 \left[ \frac{\cosh^3(x)}{30} \left(3 \cosh(2x) - 7\right) +
 \frac{4}{30}  \right] \ .
\end{align}
Minimizing the above action for $x$ with fixed ${\cal F}$ and then plugging in the expression for ${\cal F}$ yields an equation which determines $x$
\begin{align}
L^5\left(1-6 q^2 \cosh^2(x) + 5 \cosh(2x)\right) + e^{\phi_0} \sqrt{1-q^2} Q_6 \sinh(2x) = 0
\ .
\end{align}
Again, this equation can also be obtained using the general equations derived in \cite{Skenderis:2002vf}.  Plugging in the explicit expressions for $Q_6$, $\phi_0$ and $L$ leads to
\begin{align}
-1 + 3q^2 +(3 q^2 -5) \cosh(2x) + 5 \sqrt{1-q^2} \sinh(2x) = 0 \ .
\end{align}
This can be solved to give $q$ in terms of $x$.  There are two solutions
\begin{align}
\label{qsolsymm}
&q = \pm \frac{1}{\cosh(x)} \ ,&
&q = \pm \sqrt{\frac{13+5 \cosh(2x)}{9 + 9 \cosh(2x)}} \ .&
\end{align}
As we will see, the first solution is compatible with supersymmetry while the second is not.
The quantization condition is given in Appendix \eqref{F1chargesymm}
\begin{align}
N_{\rm F1} = \frac{9}{4} n N \frac{q}{\sqrt{1-q^2}} \sinh^3(x) \ ,
\end{align}
with $N_{\rm F1}$ the number of fundamental strings dissolved into the
D4-brane.  After plugging in for $x$, we obtain an expression giving
$N_{\rm F1}$ in terms of $x$
\begin{align}
N_{\rm F1}^{(1)} =& \, n N \frac{9}{4} \sinh^2(x) \ .
\end{align}
Note that in this case, $N_{\rm F1}^{(1)}$ is unbounded as $x \rightarrow \infty$ while it goes to zero for $x \rightarrow 0$.  This is consistent with the matching of these D4-brane embeddings to symmetric representations.
Computing the on-shell action gives
\begin{align}
S_{\rm D4} =& - T_4 \int d^5\xi \, e^{-\phi} \sqrt{ -\det(G_{ij} + {\cal
 F}_{ij})} + T_4 \int \hat C_{(5)} - \frac{N_{\rm F1}}{2 \pi l_s^2} \int {\cal F}
\cr
=& \frac{3 n N}{2} \left[ \left(1 + \frac{4}{9} \frac{N_{\rm F1}}{n N}
 \right)^\frac{3}{2} - 1 \right] S_{\rm F1} \ .
\label{SD4symm}
\end{align}
Again here we remark that in the limit of small $N_{\rm F1}$ the position of the D4-brane goes to $x=0$ where the spacetime $S^3$ vanishes and we recover the fundamental string wrapped on $AdS_2$, sitting at $(x,\alpha) = (0,\pi/2)$ as expected.

The result \eqref{SD4symm} gives the advertised result for symmetric representations \eqref{agreements}, for $n=1$,
and \eqref{quiversymmresult}, for $n>1$, after we identify $N_{\rm F1} = k_{\rm tot}$.
In the latter case the fractional D6-branes wrapping different cycles recombine into a
D4-brane, whose fundamental string charge is the sum of that of all the D6-branes.

We now check the supersymmetry of this embedding.  The projection matrix \eqref{projmat} reduces to
\begin{align}
\Gamma =& -\frac{1}{\sqrt{1-q^2}} \Gamma^{56789} - \frac{q}{\sqrt{1-q^2}} \Gamma^{234} \cr
=& \frac{1}{\sqrt{1-q^2}} \left( \id_2 \otimes \id_2 \otimes \sigma^3
 \otimes \id_2 \otimes \id_2 \right)  - \frac{q}{\sqrt{1-q^2}} \left(
 \id_2 \otimes \id_2 \otimes i \sigma^2 \otimes \sigma^1 \otimes \id_2
 \right) \ .
\end{align}
At $\alpha=\pi/2$, the projection condition \eqref{susyproj} on $\zeta_{+,+}$ reduces to $\sigma^1 \zeta_{+,+} = \zeta_{+,+}$.  The constraint $\epsilon = \Gamma \epsilon$ then reduces to
\begin{align}
\tilde \zeta_{\eta,-\eta} = \frac{1}{q} \sigma^1 \tilde
 \zeta_{\eta,-\eta} + \frac{\sqrt{1-q^2}}{q} i \sigma_2  \tilde
 \zeta_{\eta,-\eta} \ .
\end{align}
This is compatible with \eqref{susyproj}, provided we take $q = \eta / \cosh(x)$.  One can easily check that taking the second solution in \eqref{qsolanti} yields a projection on $\zeta_{+,+}$ which is incompatible with \eqref{susyproj} and thus breaks all supersymmetries.  Since the solution picks a specific sign choice for $\eta$, the D4-brane preserves half of the supersymmetries.

\subsection{Free Energy}\label{sec:Fgravity}

In \cite{Jafferis:2012iv} the authors computed the free energy on the
gravity side using holographic entanglement entropy and
obtained
\begin{align}
\label{FCFT2}
F_{\rm CFT} &= -\frac{9 \sqrt{2}}{5 \sqrt{8-N_f}} \pi \ n^{3/2} N^{5/2} \  + {\cal O}(N^{5/2}) \  .
\end{align}
To complete the picture we reproduce their result by a direct computation of the gravity action, regularized appropriately.
We follow the same method as in \cite{Assel:2012cp}.  First we truncate the IIA supergravity background to pure gravity on $AdS_6$ and then regularize the $AdS_6$ infinite volume by holographic renormalization techniques \cite{deHaro:2000xn,Emparan:1999pm,Marino:2011nm}.
This is a consistent truncation since we can replace $AdS_6$ space with any space which obeys the same Einstein equations.

In this computation we are using the supergravity background described in the last subsection.  This background contains both an orbifold singularity at $\alpha = 0$ and an orientifold singularity at $\alpha = \pi/2$.  Therefore  the supergravity description breaks down in these regions and so a priori our computation might miss an important contribution. Nevertheless we assume the correction to our result is subdominant in the large $N$ limit and the match with the gauge theory computation will justify a posteriori this assumption.

The effective action after the reduction reads
\beq
S_{\rm eff}=-\frac{1}{2\kappa_0^2}L^8 \, \int_{S^4/\bZ_n} e^{-2 \phi_0} \Big( \frac{4}{9}\Big)^2  (\sin\alpha)^{4/3} (\cos\alpha)^{3} \int_{AdS_6}
\sqrt{g_{(6)}}\,(R_{(6)}-2 \Lambda_{(6)}) \ ,
\label{Seff}
\eeq
where the subscript $(6)$ shows that the metric, Ricci scalar and the
cosmological constant are
$6$-dimensional, and we have $\Lambda_{(6)}=-10$.\footnote{  For $AdS_D$ spacetimes we have $R= \frac{2D}{D-2} \Lambda$ and $\Lambda = -\frac{(D-1)(D-2)}{2}$.}
Since we want to evaluate the on-shell action, we take
$R_{(6)}=3\Lambda_{(6)}=-30$. We therefore have
\begin{align}
S_{\rm eff}  &=- \frac{  L^{2}}{ 8 \pi^5  l_s^2} \  n^2 N^2 \  (-10) \, {\rm
 vol}(AdS_6)  \ {\rm vol}_4 \ ,
\label{Seff2}
\end{align}
The factor ${\rm vol}_4$ is a volume factor of the internal space,
with the $AdS$ warp factor taken into account:
\beq
{\rm vol}_4 = \frac{{\rm vol}(S^3)}{n}  \int d\alpha \ (\sin
\alpha)^{1/3} (\cos\alpha)^{3} \ = \  \frac{9\pi^2}{10 n}  \ .
\eeq
where we used ${\rm vol}(S^3) = 2\pi^2$.
The regularized volume of $AdS_6$
is given by\footnote{The volume of $AdS_6$ is regularized by
holographic renormalization techniques, see \cite[section
5]{Marino:2011nm} for a pedagogical introduction. The gravity action
contains the bulk action plus the Gibbons-Hawking surface term.
To regularize this action one needs to add (universal) covariant
boundary counterterms making the action finite. We can extract the volume
of pure $AdS$ from the renormalized gravity action.
In our problem we choose
Poincar\'e patch for the Euclidean $AdS_6$ so that the conformal
boundary is $S^5$; in the language of \cite{Emparan:1999pm} the
coordinates are given by formula (8) with $n=5, k=1$.
Then the action can be computed using formulas (63)-(65) of \cite{Emparan:1999pm}, where $\sigma_{k,n}=\sigma_{1,5}= \pi^3 $ is the volume of the unit 5-sphere.}
\begin{align}
 {\rm vol}(AdS_6) &= -\frac{8}{15} \pi^3 \ .
\end{align}
Combining these results and \eqref{radius},
we can verify that \eqref{Seff2} reproduces \eqref{FCFT2}.
This result matches both with the gauge theory and the holographic
entanglement entropy computations, providing a non-trivial check of the
concerned holographic dualities.

\section{Discussion}\label{sec.discussion}

In this paper we have computed the large $N$ limit of the VEVs of Wilson loops for a class of 5d $\scN=1$
SCFTs, both in field theory and in the dual massive IIA supergravity background.
It is non-trivial and
surprising that we can extract exact quantitative results about
non-renormalizable gauge theories, and we hope that our computation will
serve as a prototypical example for a deeper understanding of
more general classes of non-renormalizable theories.

For, quiver theories, we have found that a complete analysis would require more information coming from the holographic background. Especially the dependence on the flavors of the different nodes is absent from the current gravity description.
 We suspect that it could be recovered by including discrete holonomies of the B-field on the 2-cycles of the orbifold background or perhaps by appropriate couplings of the D-brane worldvolume theories to the Roman's mass $F_0$ .  A related issue is to consider the generalization to backgrounds describing quiver theories with nodes of different ranks. This would correspond on the gravity side to having fractional D4-branes (D6-branes wrapped on vanishing 2-cycles). For this purpose it would be useful to construct fully-backreacted geometries (cf.\ \cite{Yamaguchi:2006te,Lunin:2006xr,D'Hoker:2007fq}).  Further investigations in this direction would certainly improve our understanding of AdS/CFT for quiver theories/orbifold backgrounds. 

There are a number of generalizations we can consider.
We can consider defects of other dimensionality, such as surface
operators,
or place the theory on 5-manifolds other than $S^5$
(cf.\ \cite{Kim:2012gu,Haghighat:2012bm}).
We could also try to extend the analysis to
5d $\scN=1$ $USp(2N)$ theories with $N_f=8$, or to 5d $\scN=2$ theories.
This will lead to quantitative understanding of 6d $(1,0)$ theory or 6d
$(2,0)$ theory on $S^5\times S^1$, and the Wilson surfaces therein.

\section*{Acknowledgments}

We would like to thank Costas Bachas, Simone Giombi and Jaume Gomis
for discussion. This work is supported by STFC grant ST/J0003533/1 (JE)
and by Princeton Center for Theoretical Science (MY). BA thanks Perimeter Institute for hospitality during a visit.
MY would like to
thank Simons Center for Geometry and Physics and Yukawa Institute for Theoretical Physics (YKIS 2012)
 for hospitality where part of this work has been performed.
The research leading to these results has received funding from the [European Union] Seventh Framework Programme [FP7-People-2010-IRSES] under grant agreement n 269217.

\appendix

\section{Representation of $USp(2N)$} \label{sec.grouptheory}

In this Appendix we summarize representation theory of the
Lie algebra $USp(2N)$ needed for the main text, especially in section
\ref{CFT.general} (see for example \cite{FultonHarris}).
The representation is similar to the case of $U(N)$ gauge groups,
but there are important differences.

An irreducible representation of $USp(2N)$ is specified by a
Young diagram with at most $N$ rows.
This is expressed as a partition $\mu=(\mu_1, \mu_2, \ldots,
\mu_N)$, satisfying $\mu_1\ge \mu_2 \ge \ldots \mu_N\ge
0$, where $\mu_i$ denotes the number of boxes of the
$i$-th row. For simplicity we often drop from the notation
those $\mu_i$'s which
are equal to zero.
For example, $\mu=(7,5,3,2,1)$ represents
\begin{equation*}
{\rm \scriptsize \yng(7,5,3,2,1)} \ .
\end{equation*}
We can also represent this by the dual partition $\nu=\mu^T$.
In the example above, we have $\nu=(5,4,3,2,2,1,1)$.

In the body of this paper we discussed $k$-th symmetric and
anti-symmetric representations, obtained by symmetrizing (or
anti-symmetrizing) the $k$-th power of the fundamental representation.
For $k$-th symmetric representation $S_k$ is an irreducible
representation,
and is described by the
Young diagram of the form (shown for $k=7$),
\begin{equation*}
{\rm \scriptsize \yng(7)} \ .
\end{equation*}
However, the $k$-th anti-symmetric representation $A_k$ is not irreducible,
and decomposes into
several irreducible components.
The component with the largest
dimension is described by
\begin{equation*}
{\rm \scriptsize \yng(1,1,1,1,1,1)} \ .
\end{equation*}

For the computation of Wilson loops we need a character of the
representation $\mu$. This is given by the ``symplectic character'' $sp_{\mu}(x)=sp_{\mu}(x_1, \ldots, x_N)$, defined by
\beq
sp_{\mu}(x):=\frac{\det_{i,j}\left(x_i^{\mu_j+n-j+1}-x_i^{-(\mu_j+n-j+1)}\right)}{\det_{i,j}\left(x_i^{n-j+1}-x_i^{-(n-j+1)}\right)}
\ .
\eeq
This is a generalization of the standard Schur function for $U(N)$ groups,
and is invariant under the action of the Weyl group $\scW$,
generated by (1) permutations of $x_i$'s and (2) inversions $x_i\to x_i^{-1}$
for some $i$.

For our purposes, it is sometimes useful
to use another expression for $sp_{\mu}(x)$, given by the
``symplectic semistandard Young tableaux'' \cite{Sundaram}.
This is defined by a filling of the Young diagram $\mu$
with the letters $1< \bar{1} < 2<\bar{2}<\ldots < n<\bar{n}$
such that:
\begin{enumerate}[label=({\arabic*})]
\item the entries are weakly increasing along rows and strictly
      increasing down the columns,
\item all entries in row $i$ are larger than or equal to $i$.
\end{enumerate}
Given such a tableaux $T$, we can define its weight $w(T)$ by
\beq
w(T)=\prod_i {x_i}^{\# (i)-\# (\bar{i})} \ .
\eeq
Then we have
\beq
sp_{\lambda}(x)=\sum_{T:\, \textrm{shape}\, \lambda} w(T) \ .
\label{spassum}
\eeq

For example, let us consider $USp(4)$.
When we have $\mu=(2)={\scriptsize \yng(2)}$,
there are $10$ symplectic semistandard Young tableaux
\begin{equation*}
{\rm \young(11), \quad\young(12),\quad \young(22),
\quad\young(1\bara),\quad \young(2\barb), \quad
\young(1\barb), \quad \young(\bara 2),
\quad \young(\bara\bara),\quad\young(\bara\barb),\quad\young(\barb\barb),}
\end{equation*}
giving
\beq
sp_{\rm \tiny \yng(2)}(x)=x_1^2+x_1 x_2+x_2^2+2+\frac{x_1}{x_2}+
\frac{x_2}{x_1}+\frac{1}{x_1^2}+\frac{1}{x_1 x_2}+\frac{1}{x_2^2} \ .
\eeq
This gives dim\, {\tiny \yng(2)}$=10$, which is consistent with fact that
{\tiny \yng(2)} is the symmetric part of
${\rm \tiny \yng(1)} \otimes {\rm \tiny \yng(1)}$.

Similarly, when we have $\mu=(1,1)={\scriptsize \yng(1,1)}$,
there are $5$ symplectic semistandard Young tableaux
\begin{equation*}
{\rm \young(1,2), \quad\young(1,\barb),\quad \young(\bara,2),
\quad\young(\bara,\barb),\quad \young(2,\barb)} \ ,
\end{equation*}
giving
\beq
sp_{\rm \tiny \yng(1,1)}(x)=x_1x_2+\frac{x_1}{x_2}+
\frac{x_2}{x_1}+\frac{1}{x_1 x_2}+1 \ .
\eeq
This gives dim\, {\tiny \yng(1,1)}$=5$. This is smaller by one
than the dimension of the
anti-symmetric part of
${\rm \tiny \yng(1)} \otimes {\rm \tiny \yng(1)}$.
In fact, anti-symmetric part of ${\rm \tiny \yng(1)} \otimes {\rm \tiny
\yng(1)}$ decomposes into {\tiny \yng(1,1)} and a singlet.

\section{$AdS_2 \times S^3$ Slicing of $AdS_6$}
\label{app:embed}

In this section, we discuss an $AdS_2$ slicing of $AdS_6$ suitable for our problem.  To do so, we embed $AdS_6$ into 7-dimensional flat space, more precisely $\mathbb{R}^{2,5}$.  The $AdS_6$ surface is described by the equation
\begin{align}
-X_{-1}^2 - X_0^2 + X_1^2 + X_2^2 + X_3^2 + X_4^2 + X_5^2 = - L^2 \ ,
\end{align}
where the $X_i$ are flat coordinates on $\mathbb{R}_{2,5}$.

We first solve the constraint as follows
\begin{align}
&X_{-1} = L \coth(\lambda) \ ,&
&X_0 = L \frac{\sin(\varphi_1) \sinh(\varphi_2)}{\sinh(\lambda)} \ ,& \cr
&X_1 = L \frac{\sin(\varphi_1) \cosh(\varphi_2)}{\sinh(\lambda)} \ ,&
&X_i = L \frac{\cos(\varphi_1)}{\sinh(\lambda)} \hat X_i \ , \qquad \qquad
 i =2,3,4,5 \ ,&
\end{align}
where $\hat X_i$ describes a unit $S^3$.  This leads to the induced metric
\begin{align}
ds^2 = \frac{L^2}{\sinh^2(\lambda)} \left( d\lambda^2 + d\varphi_1^2 - \sin^2(\varphi_1) d\varphi_2^2 + \cos(\varphi_1)^2 ds^2_{S^3} \right) \ .
\end{align}
Upon analytically continuing $\varphi_2 \rightarrow i \varphi_2$ this leads to the Euclidean metric
\begin{align}
ds_E^2 = \frac{L^2}{\sinh^2(\lambda)} \left( d\lambda^2 + ds^2_{S^5} \right) \ .
\end{align}
We will be interested in a Wilson loop which wraps a great circle in $S^5$.  This can be taken to be a string worldsheet whose boundary sits at $\varphi_1 = \pi/2$ and wraps $\varphi_2$.

For computations, this metric is not the most efficient and it will be convenient to work with an $AdS_2 \times S^3$ slicing of $AdS_6$.  This can be introduced by solving the constraints as
\begin{align}
&X_{-1} = L \coth(\rho) \cosh(x) \ ,&
&X_0 = L \frac{\sinh(\psi)}{\sinh(\rho)} \cosh(x) \ ,& \cr
&X_1 = L \frac{\cosh(\psi)}{\sinh(\rho)} \cosh(x) \ ,&
&X_i = L \sinh(x) \hat X_i \ , \qquad i =2,3,4,5 \ ,&
\end{align}
where $\hat X_i$ again describes a unit $S^3$.  The induced metric is now given by
\begin{align}
ds^2 = L^2 \left( \frac{\cosh^2(x)}{\sinh^2(\rho)} (d\rho^2 - d\psi^2) + \sinh^2(x) ds^2_{S^3} + dx^2 \right) \ .
\end{align}
The two coordinate systems are related by first identifying the two $S^3$'s and then taking
\begin{align}
&\coth(\lambda) = \coth(\rho) \cosh(x),&
&\cot(\varphi_1) = \sinh(\rho) \tanh(x),&
&\varphi_2 = \psi \ .&
\end{align}
Reaching the boundary by taking $\rho = 0$ and $x$ finite maps to the surface with $\varphi_1 = \pi/2$.  Thus taking the string to wrap $\rho$ and $\psi$ gives a string whose boundary is the great circle described above.  Going to the Euclidean by taking $\psi \rightarrow i \psi$, we see that $\psi$ has periodicity $2 \pi$.

\section{Supersymmetry of the Background}
\label{app:susy}

First we need to work out the supersymmetry of the background.  The metric \eqref{metric} is in string frame, in Einstein frame ($g_E = e^{-\phi/2} g_s$) it becomes
\begin{align}
ds_{E}^2 = L^2 e^{-\phi_0/2} (\sin \alpha)^\frac{1}{12} \left[ ds^2_{AdS_6} + \frac{4}{9} \left(d\alpha^2 + \cos^2 \alpha ds^2_{S^3/\mathbb{Z}_n} \right) \right].
\end{align}
It will be convenient to introduce the frames
\begin{align}
&e^m = L e^{-\phi_0/4} (\sin \alpha)^\frac{1}{24}  \hat e^m, & &m =
 0,...,5 \ , \cr
&e^6 = \frac{2}{3} L e^{-\phi_0/4} (\sin \alpha)^\frac{1}{24} d\alpha ,& & \cr
&e^i = \frac{2}{3} L e^{-\phi_0/4} (\sin \alpha)^\frac{1}{24}
 \cos(\alpha) \hat e^i ,& &1=7,8,9 \ ,
\end{align}
where $\hat e^m$ are unit frames on $AdS_6$ and $\hat e^i$ are unit frames on $S^3/\mathbb{Z}_n$.  We use $M$ to collectively denote the frame indices so that $M = 0,...,9$.

In IIA supergravity, the spinor satisfies a reality condition $\epsilon^* = {\cal B} \epsilon$.  The BPS equations in string frame, after setting $B_{(2)} = 0$, are given by \cite{Romans:1985tz}\footnote{We have changed conventions as follows.  We have inverted the sign of the dilaton and rescaled it by a factor of $2$, $m$ is identified with $F_{(0)}$ and all of the fluxes have been rescaled by a factor of $2$, we have also redefined $\lambda$ and $\psi$ by multiplicative constants.}
\begin{align}
\delta \lambda =& \left[ (D_M \phi) \Gamma^M + \frac{5}{4} F_{(0)}
 e^{\frac{5}{4} \phi} + \frac{1}{96} e^\frac{\phi}{4} (F_{MNPQ}
 \Gamma^{MNPQ}) \right] \epsilon = 0 \ , \\
\delta \psi_M =& \left[ D_M - \frac{1}{32} F_{(0)} e^{\frac{5}{4} \phi}
 \Gamma_M + \frac{1}{128} \frac{e^{\frac{\phi}{4}}}{2} F_{NPQR} (
 \Gamma_M {}^{NPQR} - \frac{20}{3} \delta_M {}^N \Gamma^{PQR} ) \right]
 \epsilon = 0 \ . \no
\end{align}
Plugging in the solution summarized in section \ref{sec.gravity}, the dilatino equation reduces to the projection condition
\begin{align}
\label{dilatino}
\epsilon = \left[\cos(\alpha) \Gamma^6 - \sin(\alpha)
 \Gamma^{6789}\right] \epsilon \ .
\end{align}

To reduce the gravitino equation, we introduce the $\Gamma$ matrices as
\begin{align}
&\Gamma^m = \gamma^m \otimes \sigma^1 \otimes \id_2 \ ,&
&\Gamma^6 = \id_8 \otimes \sigma^2 \otimes \id_2 \ ,&
&\Gamma^i = \id_8 \otimes \sigma^3 \otimes \gamma^i  \ ,&
\end{align}
where $\gamma^m$ satisfy $\{\gamma^m,\gamma^n\} = 2 \eta^{mn}$ and $\gamma^i$ satisfy $\{\gamma^i,\gamma^j\} = 2 \delta^{ij}$.
Introduce $\gamma^{\sharp} = i \gamma^{012345}$ and ${\cal B}_{(6)}$ and ${\cal B}_{(3)}$ by
\begin{align}
&(\gamma^m)^* = {\cal B}_{(6)} \gamma^m {\cal B}_{(6)}^{-1} \ ,&
&(\gamma^i)^* = -{\cal B}_{(3)} \gamma^i {\cal B}_{(3)}^{-1} \ ,&
\end{align}
and so that they satisfy ${\cal B}_{(6)}^* {\cal B}_{(6)} = - \id_8$ and ${\cal B}_{(3)}^* {\cal B}_{(3)} = - \id_2$.
In terms of these quantities, we can write ${\cal B}$ as ${\cal B} = {\cal B}_{(6)} \otimes \sigma^1 \otimes {\cal B}_{(3)}$ and we have $(\Gamma^M)^* = {\cal B} \Gamma^M {\cal B}^{-1}$.
Next we introduce Killing spinors $\chi^{(6)}_{\eta_1}$ and $\chi^{(3)}_{\eta_2}$ on $AdS_6$ and $S^3/\mathbb{Z}_n$ respectively, which satisfy the equations
\begin{align}
\left( \hat e_m^\mu \hat \nabla_{\mu} - \frac{\eta_1}{2} \gamma_m
 \right) \chi^{(6)}_{\eta_1} = 0 \ ,\cr
\left( \hat e_i^\mu \hat \nabla_{\mu} - i \frac{\eta_2}{2} \gamma_i
 \right) \chi^{(3)}_{\eta_2} = 0 \ .
\end{align}
Using the symmetries of the above equations, we impose the conditions
$\gamma^{\sharp} \chi^{(6)}_{\eta_1} = \chi^{(6)}_{-\eta_1}$, $\chi^{(6) *}_{\eta_1} = {\cal B}_{(6)} \chi^{(6)}_{\eta_1}$ and $\chi^{(3) *}_{\eta_2} = {\cal B}_{(3)} \chi^{(3)}_{\eta_2}$.
We next decompose $\epsilon$ in the above basis of Killing spinors
\begin{align}
\epsilon = \sum_{\eta_1,\eta_2} \chi^{(6)}_{\eta_1} \otimes
 \zeta_{\eta_1,\eta_2} \otimes \chi^{(3)}_{\eta_2} \ .
\end{align}
The reality condition reduces to
\begin{align}
\zeta^* = \sigma_1 \zeta \ ,
\end{align}
and we can express the covariant derivatives of $\epsilon$ along the symmetric spaces as
\begin{align}
D_m \epsilon =& - \sum_{\eta_1,\eta_2} \frac{\eta_1}{2}
 \frac{e^{\phi_0/4}}{L \sin^\frac{1}{24}(\alpha)} \Gamma_m \Gamma^{6789}
 \left(\chi^{(6)}_{\eta_1} \otimes \zeta_{\eta_1,\eta_2} \otimes
 \chi^{(3)}_{\eta_2}\right) + \frac{1}{2} (e_m \cdot \omega_{n6})
 \Gamma^{n6} \epsilon  \ , \\
D_i \epsilon =& - \sum_{\eta_1,\eta_2} \frac{\eta_2}{2} \frac{3}{2}
 \frac{e^{\phi_0/4}}{L \cos(\alpha) \sin^\frac{1}{24}(\alpha)} \Gamma_i
 \Gamma^{789} \left(\chi^{(6)}_{\eta_1} \otimes \zeta_{\eta_1,\eta_2}
 \otimes \chi^{(3)}_{\eta_2}\right) + \frac{1}{2} (e_i \cdot
 \omega_{j6}) \Gamma^{j6} \epsilon \ , \no
\end{align}
where $\omega_{MN}$ is the spin-connection defined by $de^M + \omega^M {}_N e^N = 0$ with
\begin{align}
&\omega^m {}_6 = \frac{e^{\phi_0/4}}{16 L} \frac{\cos(\alpha)}{\sin^\frac{25}{24}(\alpha)} e^m,&
&\omega^i {}_6 = \frac{3 e^{\phi_0/4}}{2 L \sin^\frac{1}{24}(\alpha)}
 \left( \frac{\cos(\alpha)}{24 \sin(\alpha)} -
 \frac{\sin(\alpha)}{\cos(\alpha)} \right) e^i\ .&
\end{align}
The gravitino equation along $AdS_6$ reduces to
\begin{align}
\epsilon = -16  \sum_{\eta_1,\eta_2}  \eta_1 \sin(\alpha) \Gamma^{6789}
 \left(\chi^{(6)}_{\eta_1} \otimes \zeta_{\eta_1,\eta_2} \otimes
 \chi^{(3)}_{\eta_2}\right) + \cos(\alpha) \Gamma^6 \epsilon  + 15
 \sin(\alpha) \Gamma^{6789} \epsilon \ .
\end{align}
This reduces to the dilatino equation provided $\zeta_{-,\eta_2} = 0$.  The gravitino equation along $S^3/\mathbb{Z}_n$ reduces to
\begin{align}
\epsilon = - 24  \sum_{\eta_1,\eta_2}  \eta_2 \tan(\alpha) \Gamma^{789}
 \left(\chi^{(6)}_{\eta_1} \otimes \zeta_{\eta_1,\eta_2} \otimes
 \chi^{(3)}_{\eta_2}\right) +25 \cos(\alpha) \Gamma^6 \epsilon -
 \frac{24}{\cos(\alpha)} \Gamma^6 \epsilon - 25 \sin(\alpha)
 \Gamma^{6789} \epsilon \ .
\end{align}
This reduces to the dilatino equation provided $\zeta_{\eta_1,-} = 0$ and $\zeta_{+,+}$ is the only surviving component.  Since $\chi^{(6)}_{+}$ has 8 real degrees of freedom and $\chi^{(3)}_{+}$ has 2 real degrees of freedom, we conclude that there are 16 real supersymmetries.

It will be convenient to further decompose $AdS_6$ into $AdS_2 \times S^3$ slices using the coordinates given in \eqref{AdSslice}.  We denote the directions along $AdS_2$ as $m_1 = 0,1$ and the directions along $S^3$ as $m_2 = 2,3,4$.  Introducing Killing spinors $\tilde \chi^{(2)}_{\eta_3}$ and $\tilde \chi^{(3)}_{\eta_4}$ on $AdS_2$ and $S^3$ respectively, we can write $\chi^{(6)}_{+}$ as
\begin{align}
\chi^{(6)}_{+} = \sum_{\eta_3,\eta_4} \tilde \chi^{(2)}_{\eta_3} \otimes
 \tilde \chi^{(3)}_{\eta_4} \otimes \tilde \zeta_{\eta_3,\eta_4} \ .
\end{align}
As before, we can impose reality conditions on $\tilde \chi^{(2)}_{\eta_3}$ and $\tilde \chi^{(3)}_{\eta_4}$.  The reality condition on $\chi^{(6)}_{+}$ then leads to a reality condition on $\tilde \zeta_{\eta_3,\eta_4}$.  We write the $\gamma$ matrices as
\begin{align}
&\gamma^{m_1} = \tilde \gamma^{m_1} \otimes \id_2 \otimes \sigma^1,&
&\gamma^{m_2} = \id_2 \otimes \tilde \gamma^{m_2} \otimes \sigma^2,&
&\gamma^{5} = \id_2 \otimes \id_2 \otimes \sigma^3 \ ,&
\end{align}
where the $\tilde \gamma^{m_1}$ satisfy $\{\tilde \gamma^{m_1},\tilde \gamma^{n_1}\} = 2 \eta^{{m_1}{n_1}}$ and $\tilde \gamma^{m_2}$ satisfy $\{\tilde \gamma^{m_2},\tilde \gamma^{n_2}\} = 2 \delta^{m_2 n_2}$.
Proceeding similarly as before, we arrive at the projections
\begin{align}
\tilde \zeta_{\eta_3,\eta_4} =& \eta_3 \left( i \sigma^2 \sinh(x) +
 \sigma^1 \cosh(x) \right) \tilde \zeta_{\eta_3,\eta_4} \ , \cr
\tilde \zeta_{\eta_3,\eta_4} =& - \eta_4 \left( i \sigma^2 \sinh(x) +
 \sigma^1 \cosh(x) \right) \tilde \zeta_{\eta_3,\eta_4}  \ .
\end{align}
The first equation comes from the Killing spinor equation along $AdS_2$, while the second equation comes from the Killing spinor equation along $S^3$. The compatibility of these two equations sets $\tilde \zeta_{+,+} = \tilde \zeta_{-,-} = 0$.  This leaves 8 real degrees of freedom for $\chi^{(6)}_{+}$ as expected.

\section{Supersymmetry Conditions}

The conditions for supersymmetry of the probe Dp-brane are derived in \cite{Cederwall:1996ri}.  We summarize the results here in the conventions of \cite{Skenderis:2002vf}.  A probe Dp-brane embedding preserves supersymmetries which are consistent with the projection
\begin{align}
\epsilon = \Gamma \epsilon \ ,
\end{align}
where the matrix $\Gamma$ is defined by the following equation
\begin{align}
\label{projmat}
d^{p+1}\!\xi \, \Gamma = - \frac{1}{\sqrt{ -\det(G_{ij} + {\cal F}_{ij})}}
 e^{\cal F} \wedge X|_{\rm vol} \ .
\end{align}
The quantity $X$ is a sum of world-volume $\Gamma$-matrices:
\begin{align}
X =& \bigoplus_n \left( \frac{1}{(2n+1)!} d \xi^{i_{2n+1}} \wedge ... \wedge d \xi^{i_1} \Gamma_{i_1...i_{2n+1}} \right) (\Gamma_\sharp)^{n+1} ,
\end{align}
where the $\Gamma_i$ are pullbacks of space-time $\Gamma$-matrices so that $\Gamma_{i_1...i_n} = \p_{i_1} X^{m_1}...\p_{i_n} X^{m_n} \Gamma_{m_1...m_n}$ and the chirality matrix is given by $\Gamma_\sharp = \Gamma_{0123456789}$.

\section{Quantization of World-Volume Flux}

Here, we follow closely \cite{Yamaguchi:2006tq} and
\cite{Skenderis:2002vf}.  The Dp-brane action including the coupling of
the world-volume gauge field to the boundary of a stack of $N_{\rm F1}$ fundamental strings is given by
\begin{align}
S_{Dp} =& - T_p \int d^{p+1}\xi \, e^{-\phi} \sqrt{ -\det(G_{ij} + {\cal F}_{ij})} + T_p \int e^{\cal F} \wedge \hat C
+ N_{\rm F1} \int_{\p F1} ds \cdot A \ ,
\end{align}
where $G_{ij}$ is the pullback of the space-time metric, in string frame, $\hat C$ is the pullback of the RR-forms and ${\cal F} = (2 \pi l_s^2) F + \hat B_{(2)}$, where $\hat B_{(2)}$ is the pullback of the NSNS two-form and $F$ is a world volume flux with $F = d A$.  Note that $\hat C_{(p)}$ is really defined as the gauge potential of the pullback of $F_{(p+1)}$ so that $\hat d \hat C_{(p)} = \hat F_{(p+1)}$.

We shall restrict to the case where $B_{(2)} = 0$.  It is convenient to introduce the matrix
\begin{align}
M_{ij} = (\p_i X^M \p_j X^n g_{MN} + F_{ij}) \ ,
\end{align}
where $X^M$ are coordinates on the space-time.  We also define the inverse matrix $M^{ij}$, with upper indices, and the anti-symmetric part $\theta^{ij} = (M^{ij}-M^{ji})/2$.
Varying with respect to the world-volume gauge field yields the equation
\begin{align}
\p_i \left( e^{-\phi} \sqrt{-M} \theta^{ij} \right)
- \epsilon^{j i_2...i_{p+1}} \sum_{n \geq 0} \frac{1}{n!(2!)^n(p-2n)!} ({\cal F}^n)_{i_2...i_{2n+1}} F_{i_{2n+2}...i_{p+1}}
= \frac{1}{2\pi l_s^2} \frac{N_{\rm F1}}{T_p} j_j^{(F1)} \ ,
\end{align}
where $j_i^{(F1)}$ is the fundamental-string current.\footnote{Note that the extra factor of $1/(2 \pi l_s^2)$ comes from restoring the factors of $2 \pi l_s^2$ in \cite{Skenderis:2002vf}.}
Introducing a flat metric and treating $e^{-\phi} \sqrt{-M} \theta$ as a two-form, this equation can be re-expressed as
\begin{align}
& d * (e^{-\phi} \sqrt{-M} \theta) - \sum_{n \geq 0} \frac{1}{n!} d
 [({\cal F}^n) \wedge C_{(p - 1 -2n)}] = - (-1)^p \frac{N_{\rm F1}}{2\pi
 l_s^2T_p} *j^{(F1)} \ ,
\end{align}
where we have used the fact $d {\cal F} = 0$ when $B_{(2)} = 0$.
Integrating the above equation over a $p$-volume $V_{p}$ which is orthogonal to the boundary of the fundamental string, we have
\begin{align}
& \int_{M_{p-1}} \left[ * (e^{-\phi} \sqrt{-M} \theta) - \sum_{n \geq 0}
 \frac{1}{n!} ({\cal F}^n) \wedge \hat C_{(p - 1 -2n)} \right]  =
 \frac{N_{\rm F1}}{2\pi l_s^2T_p} \ ,
\end{align}
where $M_{p-1}$ is a $p-1$-dimensional closed surface which encircles the fundamental string.

For the D4-brane of section \ref{antiD4}, which wraps the internal $S^3$, the above expression reduces to
\begin{align}
\int_{S^3/\mathbb{Z}_n} \left[ * (e^{-\phi} \sqrt{-M} \theta) - C_{(3)}
 \right]  = \frac{N_{\rm F1}}{2\pi l_s^2T_4}
\end{align}
evaluated at an arbitrary value of $\rho$ and $\psi$.
Plugging in the explicit quantities and solving for $N_{\rm F1}$ gives
\begin{align}
\label{F1chargeant}
N_{\rm F1} = N-N \frac{\sin^\frac{1}{3}(\alpha)}{6} \left(\sin(3 \alpha) + 7
 \sin(\alpha) - \frac{4q}{\sqrt{1-q^2}} \cos^3(\alpha) \right) \ .
\end{align}
For the D4-brane of section \ref{symD4}, which wraps the space-time $S^3$, the above expression reduces to
\begin{align}
\int_{S^3} \left[ * (e^{-\phi} \sqrt{-M} \theta) \right]  =
 \frac{N_{\rm F1}}{2\pi l_s^2T_4} \ ,
\end{align}
evaluated at an arbitrary value of $\rho$ and $\psi$.
Plugging in the explicit quantities and solving for $N_{\rm F1}$ gives
\begin{align}
\label{F1chargesymm}
N_{\rm F1} = \frac{9}{4} n N \frac{q}{\sqrt{1-q^2}} \sinh^3(x) \ .
\end{align}

\bibliographystyle{JHEP}
\bibliography{5d_bib}

\end{document}